\documentclass[%
 reprint,
 amsmath,amssymb,
 aps,
floatfix,
]{revtex4-2}

\usepackage{graphicx}% Include figure files
\usepackage{dcolumn}% Align table columns on decimal point
\usepackage{bm}% bold math
\usepackage{adjustbox}
\usepackage{comment}
\usepackage{placeins}
\usepackage{balance}
\usepackage{lineno}
\usepackage[hidelinks]{hyperref}% add hypertext capabilities
\usepackage{cleveref}
\renewcommand{\selectlanguage}[1]{} 

\begin{document}
%\linenumbers  %to number all lines, for correction
% \preprint{APS/123-QED}

\title{
%Cross Section Measurements of Beam-Gas Interactions and \\Ion Beam Lifetime Predictions in the CERN Proton Synchrotron
%Measurements of Charge-Changing Cross Sections in the CERN \\ Proton Synchrotron and Model Predictions from Beam-Gas Interactions
% Measurements of Beam-Gas Cross Sections in the Proton Synchrotron\\and Vacuum Requirements for Future Light Ions at CERN
% Measurements of Beam-Gas Cross Sections in the Proton Synchrotron\\and Lifetime Predictions for Future Light Ions at CERN
Beam-Gas Interactions in the CERN Proton Synchrotron:\\Cross Section Measurements and Lifetime Modelling
}

\author{E.~Waagaard}
 \altaffiliation[Also at ]{CERN}%Lines break automatically or can be forced with \\
 \email{elias.walter.waagaard@cern.ch}
\affiliation{%
 EPFL, Route Cantonale, 1015 Lausanne, Switzerland
}%

%\author{Jakob Olsen}
%\affiliation{
%Aalborg University, Fredrik Bajers Vej 7K, 9220 Aalborg Øst, Denmark}%

%\author{Fenna Ukena}
% \affiliation{
%Technische Universität Berlin, Straße des 17. Juni 135,
%10623 Berlin, Germany }%

\author{F.~Ukena}
\author{J.~Olsen}
\author{R.~Alemany Fern\'andez}
\author{J.~Somoza}
\affiliation{%
 European Organization for Nuclear Research (CERN), CH-1211 Geneva 23, Switzerland
}%
\author{G. Weber}
\affiliation{GSI Helmholtzzentrum f{\"u}r Schwerionenforschung GmbH, 64291 Darmstadt, Germany}

\date{\today}

\begin{abstract}
The acceleration of high-intensity lead (Pb) beams for injection into the Large Hadron Collider (LHC) is limited by significant losses in the preceding CERN ion injector chain. A potential but largely uncharted source of losses are charge-changing beam interactions such as electron loss or capture with residual gas molecules. These effects potentially impede future ion candidate species requested by the LHC and the CERN fixed-target experiments. To predict the cross sections of charge-changing processes and the corresponding ion beam lifetimes, we present a numerical implementation combining semi-empirical electron capture and loss formulae from previous studies. We verify this numerical model with an experimental PS benchmarking campaign, measuring various beam projectile lifetimes during interactions with two in-ring gas targets (argon, helium). The target pressure profiles are reconstructed in detail from gauge measurements and vacuum simulations. At higher injected gas pressures, where beam-gas interactions dominate, measured lifetime trends and derived cross sections converge with model predictions within reported semi-empirical uncertainties, validating the package for predicting impacts on current and future ion species.

\end{abstract}

% OLD ABSTRACT
% The acceleration of high-intensity lead (Pb) beams for the Large Hadron Collider (LHC) is limited by losses in the CERN ion injector chain, partly due to uncharted charge-changing interactions (electron loss/capture) with residual gas. To predict these effects, we present \verb|beam_gas_collisions|, a publicly accessible Python package. This tool numerically implements refined semi-empirical formulae for electron loss (building on models from Weber, Shevelko, DuBois) and established models for electron capture (Schlachter) to calculate cross sections and ion beam lifetimes. We benchmarked this model in the CERN Proton Synchrotron (PS) by measuring Pb$^{54+}$ and Mg$^{7+}$ beam lifetimes during interactions with controlled in-ring argon and helium gas targets. The target pressure profiles were reconstructed in detail from gauge measurements and vacuum simulations. At higher injected gas pressures, where beam-gas interactions dominate, measured lifetime trends and derived cross sections converge with model predictions within reported semi-empirical uncertainties, validating the package for predicting impacts on current and future ion species.

%\begin{description}
%You may use the \texttt{description} environment to structure your abstract; use the optional argument of the \verb+\item+ command to give the category of each item. 
%\end{description}

\maketitle
%\tableofcontents

\section{Introduction} \label{sec:introduction}

\begin{figure}[ht]
\centering
\includegraphics[width=\columnwidth]{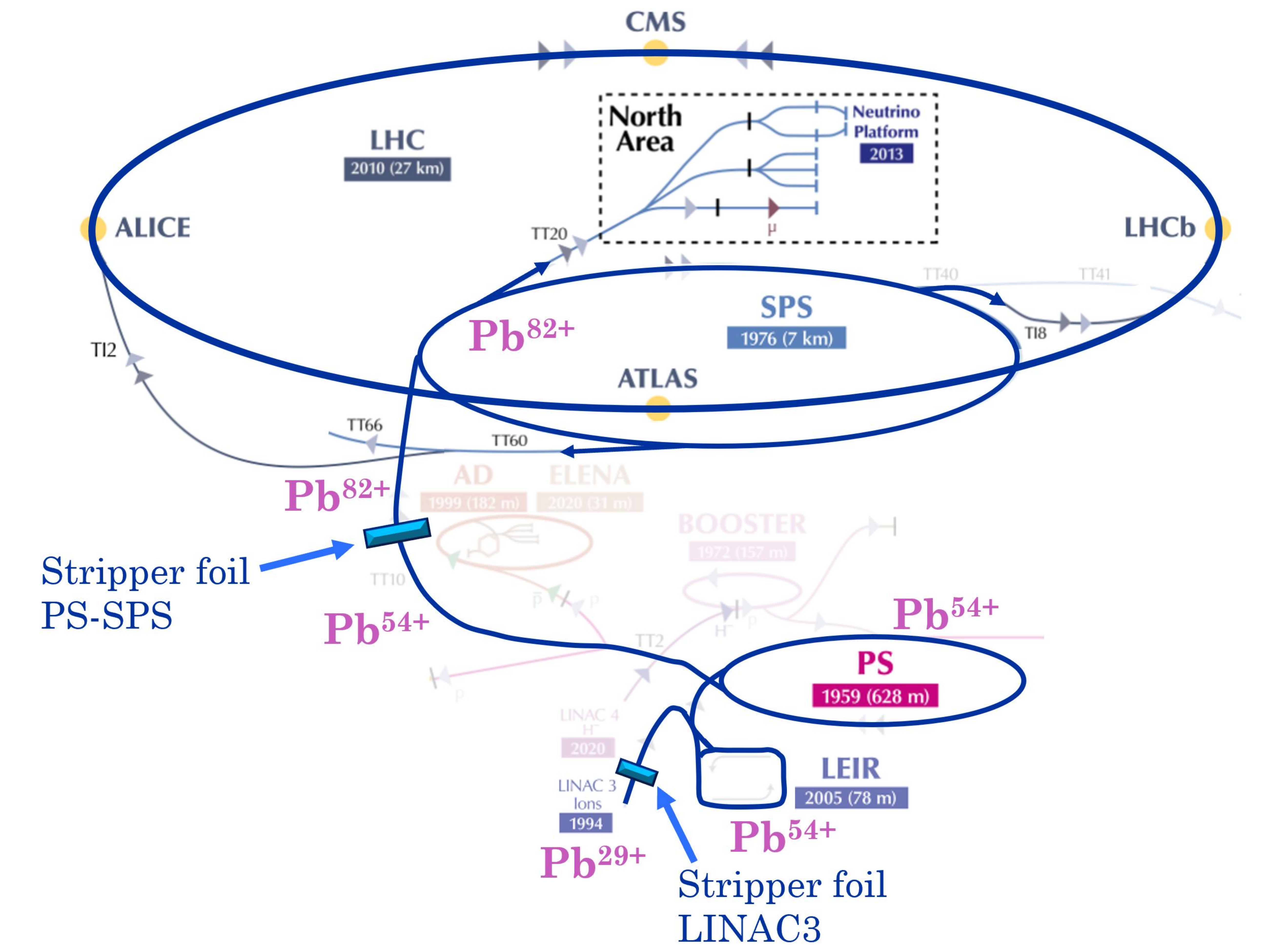}
\caption{The CERN ion injector chain, with graphics from~\cite{lopienska2022_cern_complex}.}
\label{fig:ion_accelerator_complex}
\end{figure}

The present CERN ion physics programme is based mainly on colliding lead (Pb) nuclei~\cite{benedikt_lhc_2004}. The main ion-physics users are experiments located at the Large Hadron Collider (LHC), including the ALICE detector, and fixed-target experiments at the Proton Synchrotron (PS) East Area and the Super Proton Synchrotron (SPS) North Area. The ion injector chain consists of an Electron Cyclotron Resonance (ECR) ion source, a linear accelerator (LINAC3), the Low-Energy Ion Ring (LEIR) synchrotron, the PS, and the SPS, described in detail in~\cite{j_coupard_lhc_2016}. The ion source and LINAC3 provide pulses of $^{208}\mathrm{Pb}^{29+}$ accelerated to kinetic energies of 4.2 MeV/u, which are stripped of electrons through a foil. The first stripper foil, located between LINAC3 and LEIR, strips the ion beams to $^{208}\mathrm{Pb}^{54+}$, which are then accelerated in LEIR to a kinetic energy of 72 MeV/u. PS accelerates $^{208}\mathrm{Pb}^{54+}$ ions up to 5.9 GeV/u. A second stripper foil in the PS-SPS transfer line fully strips the beam to $^{208}\mathrm{Pb}^{82+}$. The SPS provides the final acceleration of the injectors up to 177~GeV/nucleon (corresponding to $450\,Z$~GeV) before transfer to the SPS North Area or to LHC. The trajectory of lead nuclei and their charge state across the present CERN ion injector chain is shown in Fig.~\ref{fig:ion_accelerator_complex}. 
The LHC Injector Upgrade (LIU) project set the goal of doubling the total Pb beam intensity from the injector chain, leading to significant improvements in the injectors performance over several years, including momentum slip-stacking in SPS~\cite{argyropoulos_slip_stacking_2019, bartosik_injectors_2021}. Nonetheless, lead nuclei beams are still limited from significant intensity losses and emittance growth in the injectors. Space charge and intra-beam scattering (IBS) are the suspected remaining bottlenecks for the achievable ion intensity into the LHC and the North Area, studied extensively in the presence of resonances~\cite{bartosik_lhc_2017, franchetti_experiment_2010, antoniou_et_al_transverse_2017, hernandez_space_2018}.

Another important but largely unexplored beam loss mechanism is charge-changing beam interactions with residual gas molecules in the vacuum pipe, leading to intensity loss as projectile ions change charge states and fall outside the machine acceptance. These beam-gas interactions have not been studied systematically in the CERN ion injector chain, except in a few cases. A 2017 study of $^{129}$Xe$^{39+}$ ions in the SPS suggested that the Xe ion losses were driven predominantly by these beam-gas interactions~\cite{hirlaender_lifetime_2018}. The dynamic vacuum requirements of the LEIR vacuum system for required beam lifetimes due to beam-gas interactions were calculated during the technical design stage~\cite{ mahner2007_LEIR_vacuum_system}. The physics understanding of the charge-exchange processes is not only relevant for estimating loss processes in collision with the residual gas but also useful to optimize stripper targets for the highest yield of a desired charge state~\cite{arduini_stripper_foil_1996, kroeger_stripper_foil_2022}.

Other accelerator facilities have conducted studies on this topic. Heavy ion beam lifetimes due to charge-changing gas interactions have already been discussed in the context of the experimental storage ring at GSI accelerator~\cite{weber2015_total_EL_cross_section} and the future High Energy Storage Ring (HESR) at FAIR~\cite{shevelko2018lifetimes_ricode_m}. Two important reviews of charge-changing processes of ions colliding with neutral atoms are given by Refs.~\cite{basic_atomic_interactions_book_tolstikhina2018,shevelko2011_summary}.

Ion beam lifetimes and losses from beam-gas interactions are crucial to understanding for future CERN ion programmes. Ions lighter than Pb have been proposed as an alternative to achieve higher beam intensities and nucleon-nucleon luminosities in the LHC~\cite{citron_arxiv_2018}, also highlighted by the ALICE3 detector upgrade proposal~\cite{alice_collaboration_letter_2022}. In addition, fixed-target North Area (NA) experiments at CERN have requested lighter ion species such as oxygen (O), magnesium (Mg) and boron (B) beams to study the quark-gluon plasma production mechanisms~\cite{mackowiak-pawlowska_addendum_2023}. Many ion species have not been systematically tested across the CERN ion injector chain and operational experience is very limited. $^{16}$O$^{8+}$ ions will be tested during an upcoming short LHC pilot run scheduled for 2025~\cite{bruce_O_pliot_studies_2021}, having only been tested briefly during delivery to the SPS NA experiments in 1986~\cite{browzet_performance_1987_first_oxygen}. Precise estimates of these charge-changing processes are crucial to ensure that beam-gas interactions do not hinder the operation of lighter ion species under the present vacuum conditions in the CERN accelerator complex. 

In this study, we focus on the two dominant charge-changing processes in the CERN ion injectors' energy range: electron capture and loss. Semi-empirical formulae from previous studies are presented in Section~\ref{sec:atomic_beam_gas_losses}, followed by a description of the combined numerical implementation in Section~\ref{sec:semi_empirical_formula} and how it can be utilized for ion beam lifetime predictions in Section~\ref{sec:ion_beam_lifetime}. The experimental set-up and pressure profile reconstruction efforts are discussed in Section~\ref{sec:gas_inj_in_PS}. Section~\ref{sec:experimental_results} presents measurements of beam lifetime and cross sections in the PS during a controlled pressure bump creation, compared to the model predictions.

\section{Atomic beam-gas losses} \label{sec:atomic_beam_gas_losses}

Among the charge-changing processes at the energies present in the CERN injector chain, the two dominant contributing effects to beam-gas-induced losses are 
\begin{enumerate}
    \item Electron capture (EC) by the projectile: 
    \begin{equation}
    \label{eq:EC_eq}
    X^{q+} + A \quad \rightarrow \quad X^{(q-k)+} + A^{k+}, \quad k \geq 1
    \end{equation}
    \item  Electron loss (EL) --- also denoted stripping or ionization --- by the projectile: 
    \begin{equation}
    \label{eq:EL_eq}
    X^{q+} + A \quad \rightarrow \quad X^{(q+m)+} + \Sigma A +m e^-, \quad m \geq 1
    \end{equation}
\end{enumerate}
where $X^{q+}$ denotes the incident projectile ion with charge $q$ and $A$ denotes the target atom. $\Sigma A$ represents the outgoing target, which can be ionized or excited. Both electron loss and electron capture can involve multiple capture and loss processes with many electrons. Typically each change of the charge state results in the loss of the projectile ion in an accelerator. In this study, we focus on the total EL and total EC cross sections, not differentiating between the number of lost or captured electrons. EC is the dominant process at low collision energies, but its cross section decreases rapidly at higher projectile energies due to a velocity mismatch with the orbital electrons~\cite{basic_atomic_interactions_book_tolstikhina2018}. As a general guideline, EC processes dominate for low-energy ion projectiles whereas EL becomes more important for high-energy ion projectiles. This limit depends on the relation between collision velocity and the orbital velocity of a given electron~\cite{Franzke_1992_interaction}. Systematic measurements of single- and multiple-electron EC and EL cross sections with heavy ions remain limited and often contain high uncertainties~\cite{shevelko2011_summary}.

Other loss mechanisms due to residual gas include inelastic nuclear interactions and multiple elastic nuclear Coulomb scattering, but have typically lower cross sections compared to EC and EL in these energy ranges. Both these processes were previously discarded as negligible in the HERA proton ring~\cite{zimmermann1993emittance_HERA} and for protons in RHIC~\cite{rhic_proton_lifetimes_luo2016}. Multiple scattering is even more improbable for heavier ions compared to protons~\cite{Franzke_1992_interaction}, and will not be considered in this study.

\subsection{Electron capture} \label{sec:EC}

There exists a wide range of theoretical models to estimate the EC cross section, including the Classical-Trajectory Monte Carlo (CTMC) method~\cite{olson1988electronic_EC_classical_trajector_MC}, the eikonal approximation~\cite{eichler1981_eikonal}, the Coulomb Distorted Wave (CDW) approach and the normalized Brinkman-Kramers (NBK) approximation~\cite{shevelko2003target_capture_code}. These theoretical approaches constitute pillars for several available simulation codes, which only address non-radiative electron capture (NRC) since it is dominant compared to radiative electron capture (REC) below $E = 200$ MeV/u~\cite{basic_atomic_interactions_book_tolstikhina2018}. For heavy ions colliding with individual atoms, the \texttt{CAPTURE} code~\cite{shevelko2003target_capture_code} code has proven reliable across the entire energy range. 

In a more accessible form, single-electron capture can be modelled with the Schlachter empirical scaling rule valid for a wide range of fast, highly charged projectiles in gas targets~\cite{schlachter_electron_1983}, by using  reduced coordinates $\tilde{\sigma}$ and $\tilde{E}$, generalized to target proton number $Z_T$, and charge state $q$
\begin{align}
\begin{split}
\label{eq:schlachter_tilde}
\tilde{\sigma} = & \frac{1.1 \times 10^{-8}}{\tilde{E}^{4.8}} \big[ 1 - \exp(-0.037 \tilde{E}^{2.2}) \big]
\\ & \times \big[ 1 - \exp(-2.44 \times 10^{-5} \tilde{E}^{2.6}) \big] 
\end{split}
\end{align}
where
\begin{equation}
    \tilde{E} = \frac{E}{Z_T^{1.25} q^{0.7}}.
\end{equation}
and the EC cross section is then found as,
\begin{equation}
    \label{eq:schlachter}
\sigma_{\textrm{Schl}} = \tilde{\sigma}\frac{q^{0.5}}{Z_T^{1.8}},
\end{equation}
for $q \geq 3$ and $\tilde{E} \geq 10$. $Z_T$ represents the target nuclear charge, $E$ is the projectile kinetic energy in keV/u, and $\sigma$ is the EC cross section in cm$^2$/atom. For $q < 3$, the charge must be replaced by $q+0.4$. At high reduced energies above $\tilde{E} > 1000$, the EC cross section approaches
\begin{equation}
\label{eq:schlachter_high_energies}
\sigma_{\textrm{Schl}} = 1.1\times 10^{-8} \text{ [cm$^2$]} \times \frac{ q^{3.9} Z_T^{4.2} }{E^{4.8}}.
\end{equation}
Present theoretical models succeed in describing experimental $\sigma_{\textrm{EC}}$ data within a factor 2-3, but this discrepancy between theory and experiment can reach an order of magnitude in some cases and should be used with caution at lower energies~\cite{shevelko2011_summary}. Good overall agreement has been found between the Schlachter formula in Eq.~\eqref{eq:schlachter} and the \texttt{CAPTURE} code, in particular above 1-3 MeV/u~\cite{shevelko_electron_2010}.

\subsection{Electron loss} \label{sec:EL}

The theoretical framework of ionizing electron loss has been developed from the relativistic Born approximation while considering the magnetic interactions between the colliding particles, treating arbitrary projectiles and target particles with arbitrary quantum states of projectile electrons~\cite{baur2009ionization_EL_data_and_Born_approx}. These theoretical formulae constitute the foundations for several simulation codes: \texttt{LOSS} code used for non-relativistic energies \cite{loss}, and the \texttt{LOSS-R} used for relativistic energies \cite{loss-r}, and \texttt{RICODE} to calculate single-electron loss cross sections in ion-atom collisions~\cite{tolstikhina2014influence_el_RICODE}. \texttt{RICODE} includes the relativistic interaction between colliding particles but uses non-relativistic radial wave functions, assuming that the main contribution to the cross section --- ionization of outer-shell electrons --- can be treated as non-relativistic one for heavy many-electron projectiles~\cite{shevelko2011_summary}. The subsequent \texttt{RICODE-M} corrects for the relativistic factor by finding the relativistic radial functions solving either the Dirac-Fock radial equations or the Schrödinger equations based on the value of the binding energies~\cite{tolstikhina2014influence_ricode_m_first_suggested}. 

A semi-empirical formula has been mentioned by Shevelko~\cite{shevelko2011_summary} as a method to estimate the electron loss cross sections over a wide energy range based on properties of the Born approximation and numerical results from the \texttt{RICODE} program for heavy many-electron projectiles:
\begin{align}
\begin{split}
\label{eq:shevelko_semi_empirical_EL}
\sigma_{\mathrm{Shevelko}} = & 0.88\cdot 10^{-16} \textrm{ [cm$^2$/atom]} (Z_T + 1)^2 \\ & \times \frac{u}{u^2+3.5}\bigg(\frac{\textrm{Ry}}{I_p}\bigg)^{1+0.01q}\bigg(4 + \frac{1.31}{n_0}\ln(4u+1)\bigg), \\
\end{split}
\end{align}
where scaled ion projectile energy $u$ is
\begin{equation}
\label{eq:semiempirical_u_velocity}
u =\frac{v^2}{I_p/\textrm{Ry}} = \frac{(\beta c)^2}{I_p/\textrm{Ry}},
\end{equation}
and $v$ denotes the projectile velocity, $q$ the charge, $I_p$ is the projectile ionization potential in Ry units, 1 Ry = 13.606 eV. The principal quantum $n_0$ stands for number of the outermost populated electron shell of the projectile, which for ions in the ground state is tabulated together with the ionization energy in the NIST database~\cite{nist_database}. The accuracy of the semi-empirical formula in Eq.~\eqref{eq:shevelko_semi_empirical_EL} is stated to be within a factor 2 at scaled energies $u > 2$ as compared to results from \texttt{RICODE} simulations. 

A study by DuBois et al. in 2011 attempted to find a target-scaling dependence law for the EL cross section of heavy ion projectiles, considering Born scaling and contributions from screening, anti-screening and relativistic effects~\cite{dubois_electron_2011}. The model estimates the EL cross section per target atom from any projectile colliding with any target to scale as 
\begin{equation}
    \sigma_{\mathrm{Dubois}} = \sigma_{H}\bigg[N_{T_{\textrm{eff}}} + \left(Z_{T_{\textrm{eff}}} \cdot \exp\bigg(\frac{-Z_{T_{\textrm{eff}}}}{v}\bigg)\right)^2\bigg],
    \label{eq:dubois_EL}
\end{equation}
where $\sigma_{H}$ is the stripping cross section via interactions with atomic hydrogen, $v$ is the impact velocity in atomic units, $N_{T_{\textrm{eff}}}$ and $Z_{T_{\textrm{eff}}}$ are the effective charges seen by the projectile electron for the screening and anti-screening channels, respectively:
\begin{align}
    Z_{T_{\textrm{eff}}} & = Z_T\Big[1-\sqrt{1-\frac{N_{Teff}}{Z_T}}\Big], \\
    N_{T_{\textrm{eff}}} & = \langle n\rangle + A (Z_T-\langle n\rangle ) \bigg(-1+\frac{1}{\sqrt{1-\beta^2}}\bigg),
\end{align}
where $\langle n\rangle$ is the average number of active target electrons in the anti-screening channel, from Eq.~(19) in~\cite{dubois_electron_2011}. The quantity $A$ is a free parameter which in the same study was found to be approximately 10 by trial and error to provide the best fit to the experimental data. The model succeeds in most cases in predicting experimental EL cross sections for atomic or molecular targets with projectile energies ranging from a few to many tens of MeV/u, but not in all cases --- the maximum disagreement occurs at the highest energies. Hence, a different semi-empirical formula is required for energies higher than a few MeV/u present in the CERN accelerators PS and SPS.

\section{Combined semi-empirical formula for electron loss} \label{sec:semi_empirical_formula}
% for one-electron loss cross sections

\begin{figure}[t]
    \centering
    \includegraphics[width=\columnwidth]{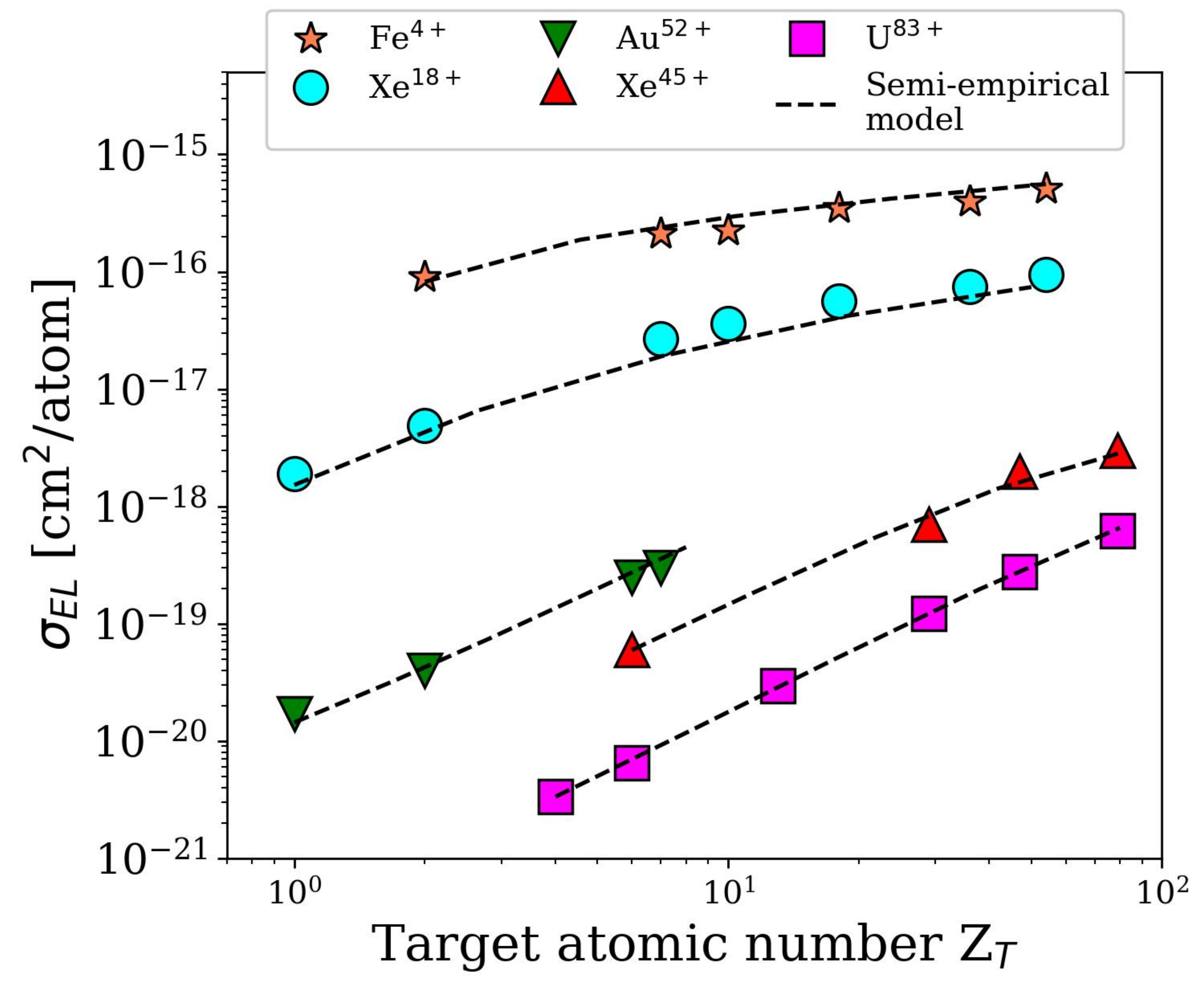}
    \caption{Semi-empirical EL model from Weber~\cite{weber_2016_semi_empirical_formula}, combining the models from Shevelko~\cite{shevelko2011_summary} and DuBois~\cite{dubois_electron_2011}, as compared to experimental data (colored points): Au$^{52+}$ at 100 MeV/u~\cite{hulskotter1991electron_Au52_EL_benchmark}, Fe$^{4+}$ at 0.36 MeV/u, Xe$^{18+}$ at 6 MeV/u~\cite{peng2004dependence_Xe18_EL_benchmark}, Xe$^{45+}$ at 140 MeV/u~\cite{meyerhof1987atomic_Xe45_EL_benchmark} and U$^{83+}$ at 955 MeV/u~\cite{meyerhof1987multiple_U83_EL_benchmark}.} % ~\cite{alton1981single_Fe4_EL_benchmark} for 20 MeV Fe, not 0.36
    \label{fig:EL_new_combined_semi_empirical_formula}
\end{figure}

The semi-empirical DuBois target-scaling model in Eq.~\eqref{eq:dubois_EL} is valid at lower energy ranges up to 70 MeV/u. Additionally,  Shevelko's formula in Eq.~\eqref{eq:shevelko_semi_empirical_EL} is not based on the \texttt{RICODE-M} version with relativistic corrections, and complementary simulated corrections are suggested beyond 100 MeV/u. Both formulae provide reasonable approximations of the electron loss cross sections, but only up to these limited energy ranges. As an alternative, Weber proposed in Ref.~\cite{weber_2016_semi_empirical_formula} a re-interpretation,  combining the treatment of the energy dependence of total electron loss by Shevelko~\cite{shevelko2011_summary} and the target-$Z$ dependence scaling formula for the projectile electron loss developed by DuBois~\cite{dubois_electron_2011} into a product of the adjusted cross sections $\tilde{\sigma}_{\textrm{DuBois}}$ and $\tilde{\sigma}_{\textrm{Shevelko}}$, also to cover higher energy ranges. This alternative semi-empirical formula gives the total electron loss cross section of many-electron ions penetrating through matter. The fitting parameters of this re-interpreted combined formula were adjusted according to more than 100 experimental data points collected over four decades with at least three different target materials over a wide energy range up to 1 GeV/u~\cite{weber_2016_semi_empirical_formula}. The single- or multi-electron loss cross section $\sigma_{\textrm{EL}}$ was reformulated by Weber as 
\begin{equation}
\label{eq:new_EL_semiempirical}
    \sigma_{\textrm{EL}} = \tilde{\sigma}_{\textrm{Shevelko}} \times \tilde{\sigma}_{\textrm{DuBois}},
\end{equation}
where
\begin{align}
\begin{split}
    \label{eq:shevelko_in_semi_empirical}
    \tilde{\sigma}_{\textrm{Shevelko}} = & \frac{c_5 (10^{-16})u}{u^2 + c_6} \left(\frac{\textrm{Ry}}{I_p}\right)^{c_7 + [1- \exp ( -c_9u^2 ) ]\frac{q+c_2}{Z_p}} \\ & \times \Big(1 + \frac{c_8}{n_0} \log{\big[(u+1)\gamma\big]}\Big), 
\end{split}
\end{align}
and 
\begin{align}
\begin{split}
    \tilde{\sigma}_{\textrm{DuBois}} &= F_1 + \Big(F_2 e^{F_3}\Big)^2, \\
\end{split}
\end{align}
with the projectile atomic number $Z_p$, the relativistic Lorentz factor 
\begin{equation}
\label{eq:Lorentz}
\gamma = \frac{1}{\sqrt{1-\beta^2}}, 
\end{equation}
and the parameter $u$
\begin{equation}
\label{eq:u_param}
    u = \frac{(\beta / \alpha)^2}{I_p/\textrm{Ry}}. 
\end{equation}
The parameters $F_1$, $F_2$ and $F_3$ are given as
\begin{align}
    F_1 &= \min \{|N_{\textrm{eff}} + c_0 (Z_T-N_{\textrm{eff}}) (\gamma - 1)|, Z_T\}, \\
    F_2 &= Z_T (1-\sqrt{1-F_1/Z_T}), \\
    F_3 &= \frac{-(F_2^{c_1})}{\left(\sqrt{2E_{\textrm{kin}}/AU} + \sqrt{2I_p/m_e}\right) / \alpha},
\end{align}
where 
\begin{equation}
\label{eq:N_eff}
N_{\textrm{eff}} = \min \{10^{c_3\log_{10}{Z_T}+c_4(\log_{10}{Z_T})^2}, Z_T \},    
\end{equation}
and  $\alpha=7.3\cdot 10^{-3}$, $AU=931$ MeV, $\textrm{Ry}=13.606$ eV, $m_e=510.998$ keV. Equation~\eqref{eq:new_EL_semiempirical} is based on experimental data from projectiles impacting on different target atomic number $Z_T$ at various energies, also highly relativistic: Au$^{52+}$ at 100 MeV/u~\cite{hulskotter1991electron_Au52_EL_benchmark}, Fe$^{4+}$ at 20 MeV/u~\cite{alton1981single_Fe4_EL_benchmark}, Xe$^{18+}$ at 6 MeV/u~\cite{peng2004dependence_Xe18_EL_benchmark}, Xe$^{45+}$ at 140 MeV/u~\cite{meyerhof1987atomic_Xe45_EL_benchmark} and U$^{83+}$ at 955 MeV/u~\cite{meyerhof1987multiple_U83_EL_benchmark}. Figure~\ref{fig:EL_new_combined_semi_empirical_formula} shows the agreement of the combined semi-empirical cross section formula in Eq.~\eqref{eq:new_EL_semiempirical} versus these experimental values. These energies, $Z_p$ and $Z_T$ values are considered a validated regime in which this EL formula can be applied.
\begin{table}[t]
%\begin{adjustbox}{width=\columnwidth}
\begin{ruledtabular}
\begin{tabular}{l|llllllllll}
Constant & $c_0$ & $c_1$ & $c_2$ & $c_3$  & $c_4$   & $c_5$   & $c_6$   & $c_7$   & $c_8$   & $c_9$   \\
Value     & 10.88 & 0.95  & 2.5   & 1.11 & -0.18 & 2.65 & 1.36 & 0.81 & 1.01 & 6.14
\end{tabular}
\end{ruledtabular}
\caption{Fitting parameters $c_i$ for Eq.~\eqref{eq:new_EL_semiempirical}~\cite{weber_2016_semi_empirical_formula}.}
\label{tab:fitting_coefficients_semiempirical}
%\end{adjustbox}
\end{table}

In this study, we have numerically implemented this combined semi-empirical model in the Python package \verb|beam_gas_collisions|~\cite{beam_gas_collisions_github_repo}, combining it with the Schlachter formula in Eq.~\eqref{eq:schlachter}. This software program can be used to calculate EC and EL cross sections and ion beam lifetimes for any accelerator with known rest gas composition, pressure and projectile state in this full energy range 0.36 to 955 MeV/u. The constants $c_i$, $i\in [0, 1 \dots 9]$ used by Weber~\cite{weber_2016_semi_empirical_formula} are displayed with two decimals in Table~\ref{tab:fitting_coefficients_semiempirical}. 

\begin{figure*}[t]
\centering
\includegraphics[width=0.95\textwidth]{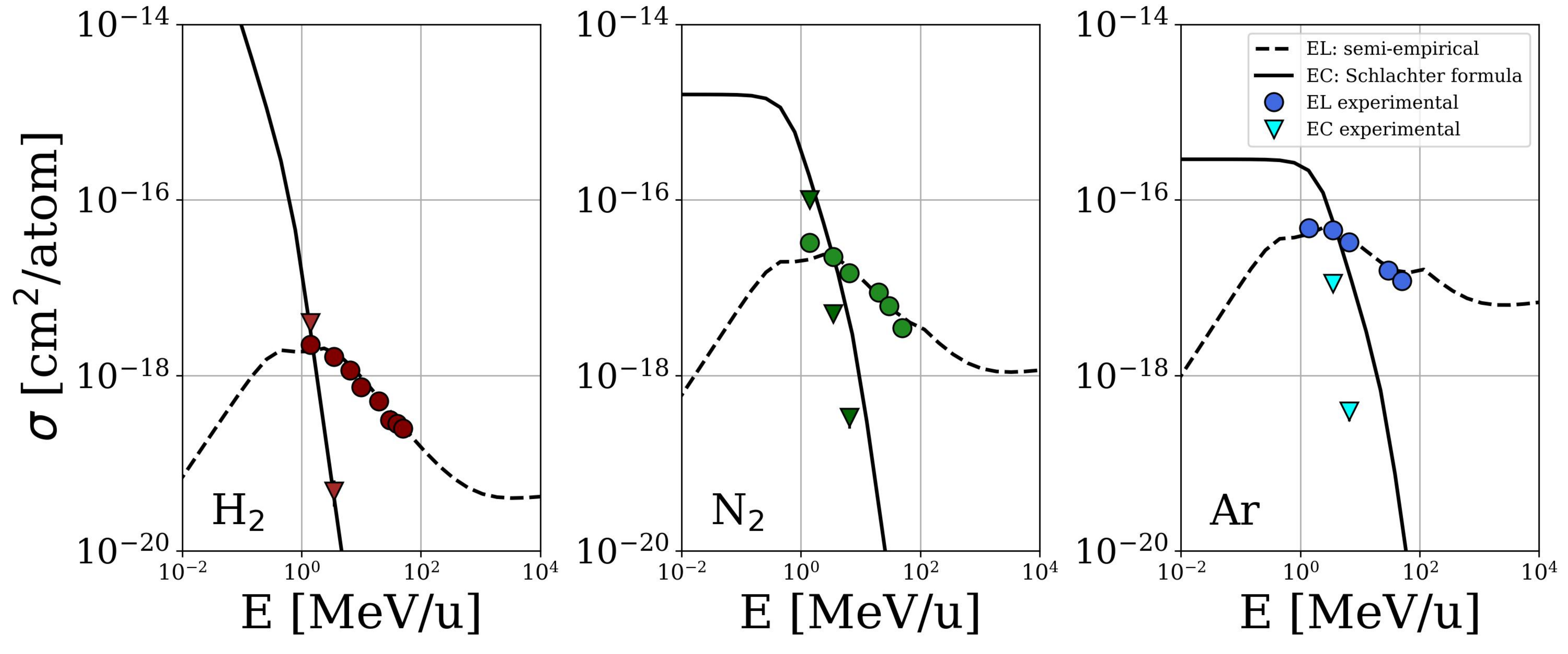}
\caption{EC and EL cross sections in collisions of U$^{28+}$ ions with H$_2$, N$_2$ and Ar targets as a function of ion energy. Solid line: EC cross section from the Schlachter formula in Eq.~\eqref{eq:schlachter}. Dashed line: EL cross section from semi-empirical model. EL experimental data points are reviewed in detail in~\cite{weber2015_total_EL_cross_section}: H$_2$~\cite{franzke1981_semi_empirical_vacuum_formula, olson2004projectile_EL, weber2015_total_EL_cross_section},  N$_2$~\cite{franzke1981_semi_empirical_vacuum_formula, olson2004projectile_EL, weber2015_total_EL_cross_section, perumal2005_multiple_EC} and Ar~\cite{erb1978_gsi_EC_data, olson2004projectile_EL, weber2015_total_EL_cross_section}. EC experimental data points: H$_2$~\cite{franzke1981_semi_empirical_vacuum_formula, olson2004projectile_EL}, N$_2$~\cite{franzke1981_semi_empirical_vacuum_formula, olson2004projectile_EL} and Ar~\cite{olson2004projectile_EL}. }
    \label{fig:EL_and_EC_vs_experiment}
\end{figure*}

An initial check of Eq.~\eqref{eq:new_EL_semiempirical} compared to additional experimental cross section data --- not used for parameter fitting ---  of U$^{28+}$ on H$_2$, N$_2$ and Ar targets over different projectile energies is shown in Fig.~\ref{fig:EL_and_EC_vs_experiment}. Although Pb is the main projectile for the CERN ion physics programmes, no such cross section data in relevant energy ranges were found. In general, one finds good agreement between the estimated EL cross sections, similar to Fig. 12 in~\cite{shevelko2011_summary}. U$^{+28}$ cross sections for Ar targets are similar on N$_2$ and lower on H$_2$ targets. Additional experimental EC data for U$^{+28}$ are shown together with Schlachter formula predictions from Eq.~\eqref{eq:schlachter} on the same targets to illustrate the energy regimes of the EC and EL. The EC cross section dominates at lower energies, before quickly decaying. On the other hand, the EL cross section forms a peak before slowly decreasing with increasing energy. The Schlachter formula agrees well with low-$Z_T$ experimental data, but some disagreement appears at around 10 MeV/u for the higher $Z_T$ target argon. For the Ar target case, the Schlachter formula above 10 MeV/u overestimates EC cross sections for Ar targets by a factor 5-10, also mentioned in~\cite{shevelko_electron_2010}. However, the EL cross section dominates by one or two orders of magnitude in this energy regime, and this EC discrepancy at energies above 10 MeV/u has a smaller impact on overall lifetime estimates in the CERN accelerators.

\section{Ion beam lifetime estimates} \label{sec:ion_beam_lifetime}  

Electron capture and loss from interactions with residual gas is highly relevant for beam lifetimes. Shevelko~\cite{shevelko2011_summary} stated that the Schlachter EC formula in Eq.~\eqref{eq:schlachter} combined with the EL cross sections formula in Eq.~\eqref{eq:shevelko_semi_empirical_EL} can be used to estimate ion beam lifetimes in accelerators~\cite{shevelko2011_summary}, but has to our knowledge not yet been systematically deployed to estimate beam-gas interactions at CERN. To highlight the relation between beam intensity loss and lifetime, we express the total ion beam intensity $I$ as a function of time $t$ as
\begin{equation}
I(t) = I(t_0)\cdot\exp\bigg(- \frac{t}{\tau}\bigg),  
\label{eq:beam_intensity}
\end{equation}
where the lifetime $\tau$ due to atomic interactions with the residual gas is
\begin{equation}
\tau = \frac{1}{\sigma\,n\,\beta \, c },
\label{eq:beam_lifetime}
\end{equation}
and $\sigma = \sigma_{\textrm{EC}} + \sigma_{\textrm{EL}}$ is the total charge-changing cross section, $\beta$ is the projectile relativistic beta factor, $c$ is the speed of light and $n$ is the molecular density in the beam pipe. Equation~\eqref{eq:beam_lifetime} has been used to calculate heavy-ion beam lifetimes in FAIR~\cite{shevelko2018lifetimes_ricode_m} and dynamic vacuum requirements of the LEIR vacuum system~\cite{mahner2007_LEIR_vacuum_system} at CERN due to beam-gas interactions. In this study, the Franzke semi-empirical formulae~\cite{franzke1981_semi_empirical_vacuum_formula} for EC and EL were used, extrapolated from results at the single projectile energy of 1.4 MeV/u and tested on He, N$_2$ and Ar targets. 

A few assumptions made in \verb|beam_gas_collisions| to treat mixtures of residual gases are worth mentioning. If several target atoms or molecules are present in the vacuum, the lifetimes from Eq.~\eqref{eq:beam_lifetime} have to be inversely added in the following manner~\cite{mahner2007_LEIR_vacuum_system}:
\begin{align}
    \label{eq:inversely_adding_lifetime}
    \frac{1}{\tau_{\text{tot}}} = \sum_{i\in \Omega}\frac{1}{\tau_{i}},
\end{align}
where $\Omega$ is the space containing all residual gases, i.e., $\Omega=\{\text{H}_2, \text{He},\ ...\}$. To estimate the molecular density $n$ for every target particle in a gas mix, we assume that the ideal gas law applies under vacuum conditions
\begin{equation}
\label{eq:particle_number_density}
n=\frac{P}{kT}, 
\end{equation}
where $P$ is the exerted pressure, $k$ is the  Boltzmann constant and $T$ is the temperature~\cite{mandl1991statistical_ideal_gas_law}, set to $T = 293$ K in these studies.

To treat the mixture of residual gases in the CERN accelerators, we estimate the partial pressure from each individual non-reacting gas in a mixture from Dalton's law~\cite{dalton1802essay}, which states that the total pressure exerted by a mixture of gases is the sum of the partial pressures of component gases: 
\begin{equation}
\label{eq:Daltons_law}
P_{\textrm{tot}} = \sum_{i=1}^n P_i =  P_{\textrm{tot}} \sum_{i=1}^n  \frac{n_i}{n_{\textrm{tot}}}
\end{equation}
where $n_i/n_{tot}$ is the molar fraction: the gas amount divided by the total amount of constituents in a gas. For the treatment of molecules, we employ the cross section rule of additivity
\begin{equation}
\label{eq:addivity_rule}
\sigma_{\textrm{mol}} = \sum_i a_i \sigma(Z_i), 
\end{equation}
where $a_i$ is the number of atoms in the molecule with atomic number $Z_i$. The validity of the additivity rule was tested with a linear relationship between the total-loss cross section and the target atomic number from noble-gas data to calculate $ \sigma(Z_i)$, and works well for electron loss from heavy ions in the MeV energy range for different charge states~\cite{additivity_rule}.

\begin{comment}
\begin{table}[t]
\begin{ruledtabular}
\begin{tabular}{c|ccc} %\hline
          & LEIR       & PS                   & SPS       \\ \hline
          & & & \\[-2.2ex]
$\bar{P}$ {[}mbar{]} & $10^{-11}$ & $1.2 \times 10^{-9}$ & $10^{-8}$ \\ \hline 
H$_2$ fraction           & 0.83       & 0.9                  & 0.905     \\
H$_2$O fraction          & 0.02       & 0.1                  & 0.035     \\
CO fraction          & 0.04       & 0                    & 0.025     \\
CH$_4$ fraction          & 0.05       & 0                    & 0.025     \\
CO$_2$ fraction          & 0.06       & 0                    & 0.01  \\ \hline 
Cycle time [s] & 3.6 & 2.4 & ~60
\end{tabular}
\end{ruledtabular}
\caption{Assumed average static pressure values $\bar{P}$ and rest gas composition (fraction of total particle density) and nominal Pb cycle time duration used in operation.}
\label{table:gas_fractions}
\end{table}
\end{comment}

\section{Ion beam lifetime experiments} \label{sec:gas_inj_in_PS}

The PS accelerator features various beam diagnostics instruments, including the Beam Gas Ionization (BGI) Profile Monitor to measure transverse beam profiles, developed with Timepix3 hybrid pixel detectors, installed 2017 in PS section 82 and successfully benchmarked against wire scanners at the same location~\cite{BGI_2016_development, BGI_2017_first}. The instrument is equipped with an external gas injection system to inject neutral gas, typically argon (Ar). The circulating beam in the accelerator interacts with the injected gas. This interaction ionizes the gas producing free electrons and positive ions. A uniform electric field is applied perpendicular to the beam direction to extract the produced electrons or ions towards a detector. From the distribution of the charged particles impacting the detector, the beam transverse profile is obtained.
%The BGI then applies external electromagnetic fields to ionize the injected gas molecules from charged particles in the beam in order to infer beam profiles from the electron distribution. 
The injection system is connected via a valve to the PS beamline and consists of a Penning gauge, the gas bottle, and a turbomolecular pump connected to the rest of the instrumentation via a valve. The gas valves open until a high enough temporal resolution is obtained for the profiles, with different programmable ``Set Points" that control the flow of neutral gas in the beam pipe. ``Set Point" can hence be perceived as the setting that roughly translates to a certain pressure. We also point out that PS titanium sublimation pumps were active before the experiments to improve base vacuum conditions.

In recent PS Machine Development (MD) studies in 2023 and 2024, the BGI-related injection system was exceptionally used to inject neutral gas into the vacuum pipe at the BGI location and generate a controlled pressure bump, effectively acting as an in-ring gas target. A special PS Ion Lifetime cycle was prepared, which stays at injection energy, coupled to either low-intensity (EARLY) or high-intensity (NOMINAL) ion beams from LEIR. By studying the ion beam intensity decay during cycles for various Set Points, one can extract the fitted beam lifetime parameter $\tau$ from Eq.~\eqref{eq:beam_lifetime}. These studies include three experimental beams with injected gas: (A) Pb$^{54+}$ on Ar, (B) Mg$^{7+}$ on Ar and (C) Pb$^{54+}$ on He. Experiment (A) was conducted in autumn 2023, (B) in spring 2024 and (C) in autumn 2024. He and Ar were the two available target gases for the BGI injection system in this period.

\subsection{PS pressure profile reconstruction}
\label{sec:pressure_profile_reconstruction}

\begin{figure*}[t]
    \centering
    \includegraphics[width=2\columnwidth]{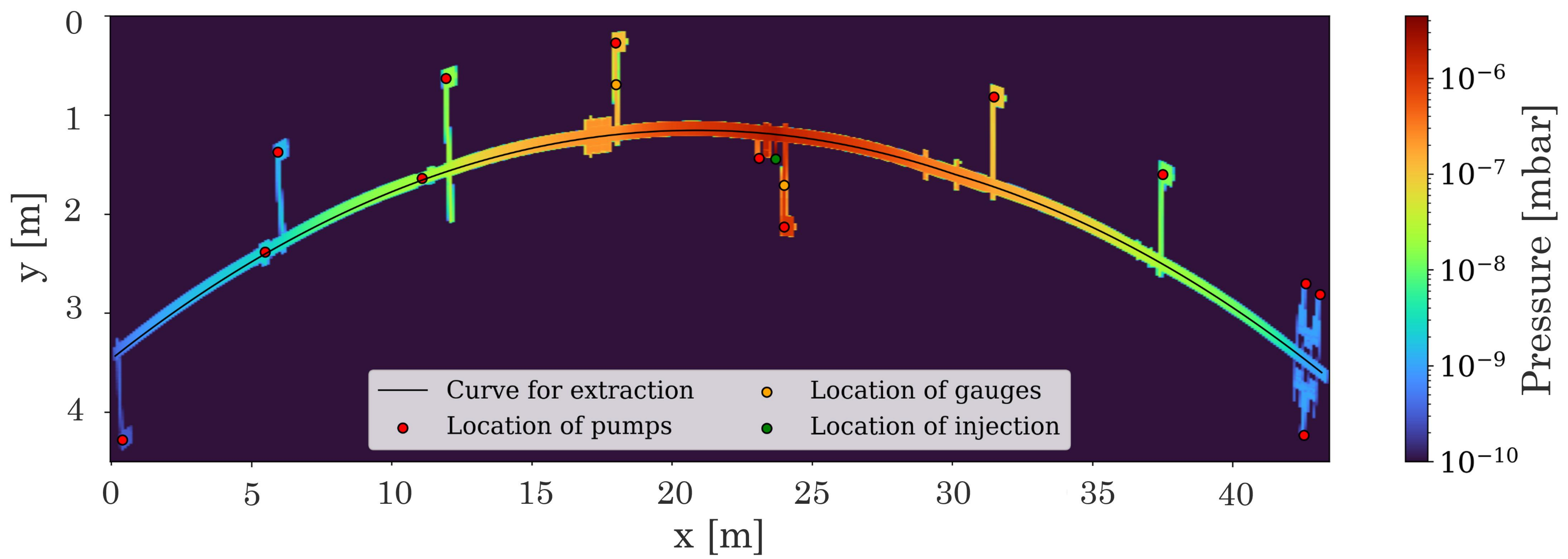}
    \caption{3D Molflow+ model image of the relevant PS sections for Set Point 160, with the simulated Ar pressure distribution (N$_2$ equivalent). Gas injection locations, pumps, and gauges are also marked.} 
    \label{fig:molfow_ps_model}
\end{figure*}

The most uncertain parameter in Eq.~\eqref{eq:beam_lifetime} is the average molecular (gas) density $n$, which requires knowledge of the PS pressure profile. The process of reconstructing the pressure profile has two main steps: measurement-based modelling and pressure simulations of gas injection experiments. During the experiment, before injecting gas into the ring, the flow was measured using an IKR070 Penning gauge while being pumped with a known turbo-pump speed. Then, the turbo was isolated and the injection valve was opened allowing this flow to create the pressure bump. The PS ring is pumped using triode ion pumps spaced approximately every 4.5 m. The measured injected flow was used as input for the pressure profile modelling. To model the pressure profile of the PS ring, a 3D representation of a significant portion of the ring ($\simeq$ 10\%) was created. This model was used to simulate the gas propagation in the ring using MolFlow+ \cite{molflowcitation, molflow_algorithm_monte_carlo_ady2016}, a Monte Carlo code for vacuum simulations. The obtained pressure profile is illustrated in Fig.~\ref{fig:molfow_ps_model}.

For the calculation of the pressure profiles, the following aspects were considered:
\begin{itemize}
    \item IKR070 Penning gauges gas correction factors
    \item IKR070 Penning gauge 30\% accuracy
    \item The pressure dependency on pumping speed of the sputter ion pumps 
    \item The saturation state of the ion pumps, and hence the observed pumping speed, evolved during the experiment as more injected gas is implanted in the surface of the ion pump \cite{agilent_vacion_300_2016,agilent_vacion_40_75_2016}
\end{itemize}

\begin{figure}[b]
    \centering
    \includegraphics[width=\columnwidth]{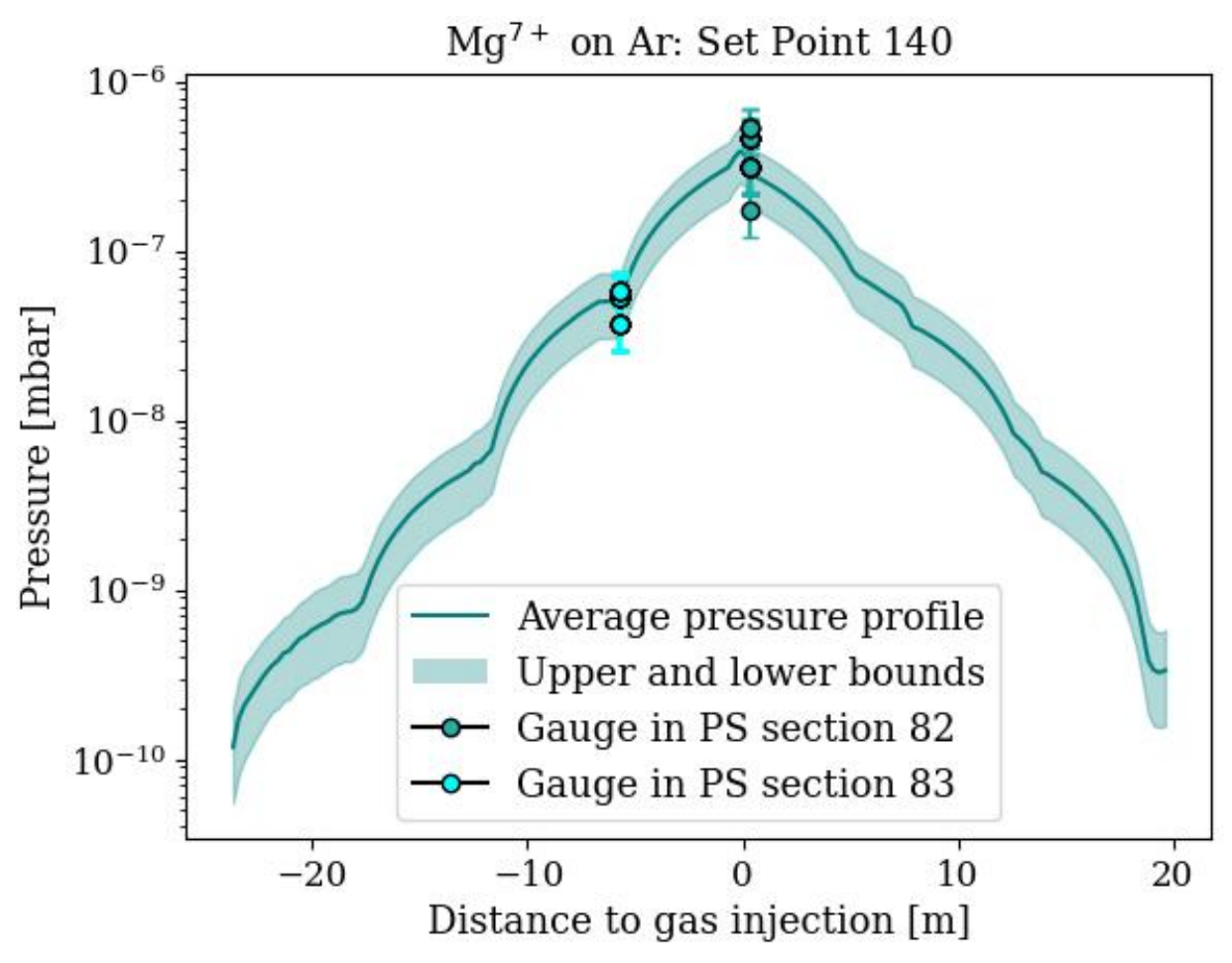}
    \caption{Example of simulated PS pressure profile with Molflow+ with injected Ar for the Mg$^{7+}$ beam test, centred around the BGI injection system. This generated pressure bump at Set Point 140 corresponds approximately to an N$_2$ equivalent average PS pressure of $5\times10^{-8}$ mbar. Additionally, two pressure measurements of two gauges near the BGI location for this configuration are shown. The lower and upper bounds of the pressure profile are determined by using the maximum and minimum pumping speed corresponding to completely unsaturated and saturated sputter ion pumps, respectively, with an additional 30\% error of the Penning gauge of the gas injection system.  }
\label{fig:Simulated_Molflow_pressures_Pb_on_Ar_2023_10}
\end{figure}

For each experimental setting, the boundaries of the pressure profile were calculated by combining the lowest possible injection according to the gauge accuracy with the unsaturated pumping speed (maximum possible pumping speed), and the highest possible flow with the saturated pumping speed (minimum). This approach was used to obtain the lower and upper pressure profile bounds respectively for each setting, which are used to compute an interval of plausible average PS total pressure. Fig.~\ref{fig:Simulated_Molflow_pressures_Pb_on_Ar_2023_10} illustrates one of these calculated profiles and its comparison with two Penning gauges in the ring. 

An important step is also to disentangle the loss effects of the controlled pressure bump from those of the baseline residual gas, which carries significant uncertainty due to e.g. outgassing of the machine and the injection system. For the highest pressure values, we assume that the controlled pressure bump around the BGI dominates the average PS pressure. For the lower pressure values, the static base pressure is most likely too big to be ignored compared to the injected gas. For this reason, we conduct a direct beam lifetime measurement before any gas injection for each experiment to find the decay constant for the loss contribution from the static base pressure and other unknown loss sources. This step is important for experimental cross section calculations. We assume additive loss rates and subtract lifetimes inversely as in Eq.~\eqref{eq:inversely_adding_lifetime} to find the experimental beam lifetime $\tau_{\textrm{bump}}$ due to controlled pressure bump
\begin{align}
\label{eq:tau_bump_lifetime_addition}
\frac{1}{\tau_{\textrm{measured}}} = \frac{1}{\tau_{\textrm{residual}}} + \frac{1}{\tau_{\textrm{bump}}} \\
\label{eq:tau_bump_lifetime_addition_2}
\Rightarrow \tau_{\textrm{bump}} = \bigg(\frac{1}{\tau_{\textrm{measured}}} - \frac{1}{\tau_{\textrm{residual}}}\bigg)^{-1}.
\end{align}
Other profile-dependent incoherent loss effects are also considered, with each experiment evaluating the impact of the longitudinal beam profile on lifetime. Tests are conducted with both RF cavity settings: ON and OFF. With the RF cavity ``ON", the beam will be bunched and other loss mechanisms such as space charge and intra-beam scattering may play a role. With the RF cavity ``OFF", the beam will be coasting and longitudinally less dense, allowing us to decouple such incoherent effects when studying the losses. Ion-induced beam losses from the pressure bump leading to outgassing from the PS vacuum chamber walls were not considered in detail here and should be addressed in future studies.

\section{Experimental Results}
\label{sec:experimental_results}

Two target gases were available in the external gas injection system to generate the in-ring target: argon (Ar) and helium (He). The projectile Pb$^{54+}$ was tested at 72.2 MeV/u for both these target gases. Additionally, a pilot beam of Mg$^{7+}$ at 90.2 MeV/u was also tested with Ar as the target gas.

\subsection{Pb\texorpdfstring{$^{54+}$}{Lg} lifetime experiment with injected Ar gas}
\label{sec:Pb54_on_Ar}

\begin{figure}[b]
\centering
\includegraphics[width=\columnwidth]{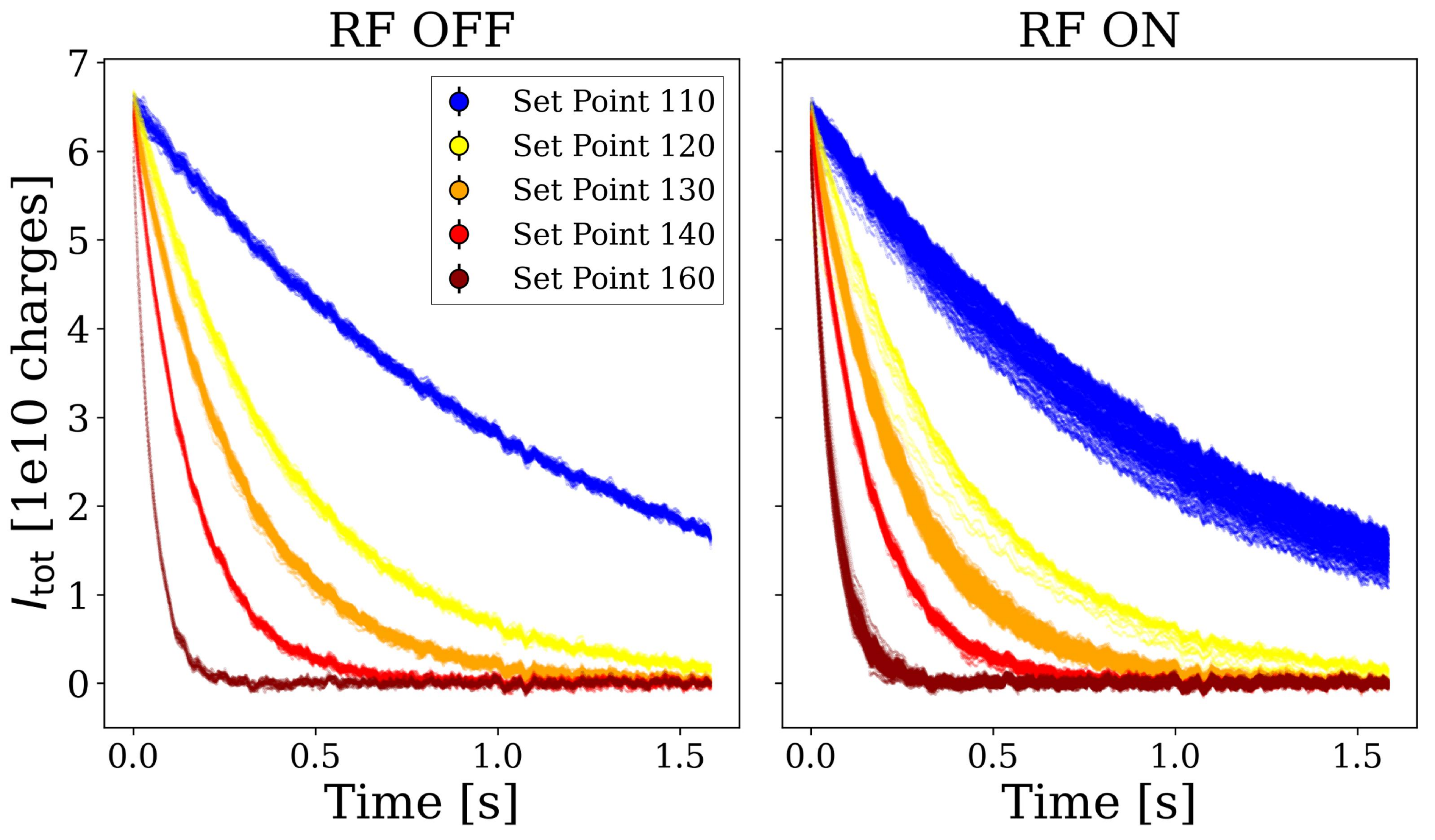}
\caption{Impact on intensity profiles for the NOMINAL Pb$^{54+}$ beam in the PS for various Set Points. Higher Set Points (red) indicate more injected target gas, and lower Set Points (blue) indicate less injected gas.}
\label{fig:ps_ar_tau_nominal_raw_data}
\end{figure}

The first conducted PS experiment (A) in  October 2023, features lifetime measurements of Pb$^{54+}$ beams at 72.2 MeV/u with injected Ar gas. Figure~\ref{fig:ps_ar_tau_nominal_raw_data} shows the Pb$^{54+}$ beam intensity profiles from the PS Beam Current Transformer (BCT) as a function of the cycle time for various SPs. Increasing the quantity of injected Ar gas causes a reduction in beam lifetime. Throughout these experiments, we also adjusted for any faulty BCT calibration by first measuring the intensity baseline without any injected beam. Different beam type permutations were tested: low-intensity EARLY and high-intensity NOMINAL beam from LEIR, bunched (RF ON), and coasting (RF OFF) beam. Figure~\ref{fig:ps_ar_tau_nominal_raw_data} clearly shows how both coasting (RF OFF, left) and bunched (RF ON, right) beams get reduced lifetimes with more injected neutral gas. The larger spread in beam lifetimes for the lowest SP (dark blue) with RF ON is explained by the higher local pressure variation during this interval before the pumping speed was saturated.

\begin{figure}[t]
\centering
\includegraphics[width=\columnwidth]{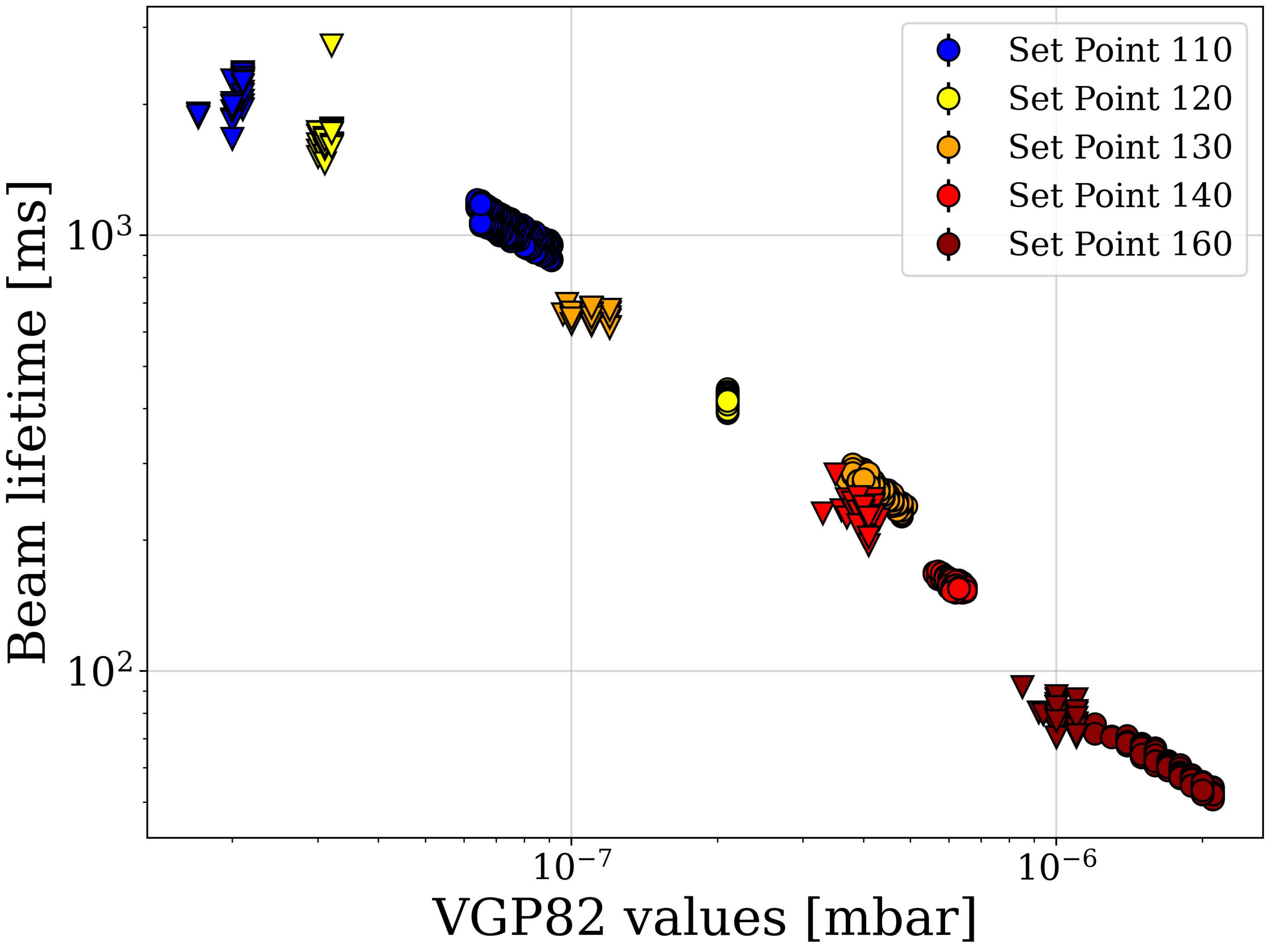}
\caption{Estimated Pb$^{54+}$ beam lifetime as a function of pressure gauge VGP82 readout. Circles represent high-intensity NOMINAL beams, triangles low-intensity EARLY beams.}
\label{fig:001_Pb54_on_Ar_PS_Lifetime_over_VGP82}
\end{figure}

In addition, the beam lifetime is monitored as a function of the measured pressure values from gauge VGP82, located close to the BGI instrumentation.  Figure~\ref{fig:001_Pb54_on_Ar_PS_Lifetime_over_VGP82} shows the fitted beam lifetime trends for various gauge readouts. Apart from small discontinuities at e.g. $10^{-7}$ mbar (most likely due to the Penning gauge switching voltage), the decaying power law trend with higher pressure readout is surprisingly uniform. For different Set Points, lifetimes of the high-intensity NOMINAL beam (circles) and low-intensity EARLY beam (triangles) seem to follow the same trend. Due to various pumping and saturation speeds discussed in Sec.~\ref{sec:pressure_profile_reconstruction}, the PS average pressure will vary slightly within the same SP, clearly shown in Fig.~\ref{fig:001_Pb54_on_Ar_PS_Lifetime_over_VGP82}. To account for this effect, the relative fluctuation in the VGP82 readout for each SP is used to scale the pressure profiles within the upper and lower bounds in Fig.~\ref{fig:Simulated_Molflow_pressures_Pb_on_Ar_2023_10} to compute the average PS pressure for each estimated beam lifetime.

\begin{figure}[t]
\centering
\includegraphics[width=\columnwidth]{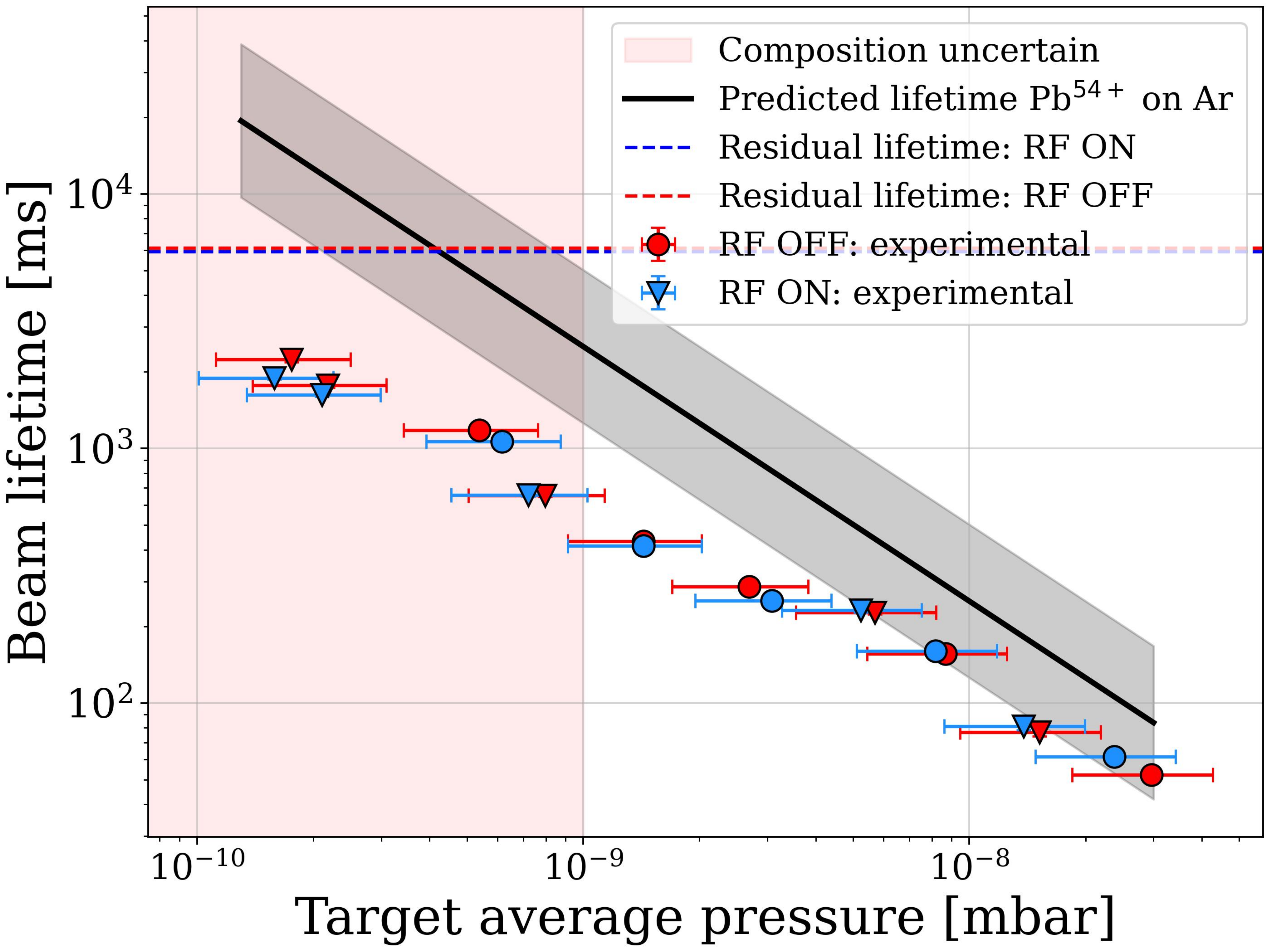}
\caption{Measured Pb$^{54+}$ beam lifetimes for estimated target average pressures with neutral Ar gas injected. The black line represents the calculated lifetime from EC and EL on Ar, and the factor 2 uncertainty is in grey. Circles represent high-intensity NOMINAL beams, triangles low-intensity EARLY beams. Red means RF OFF (coasting), and blue RF ON (bunched). Residual lifetimes without any injected gas are also included. Rest gas compositions of target pressures below $10^{-9}$ mbar are considered too uncertain.}
\label{fig:001_Pb54_on_Ar_PS_Lifetime_over_avg_Pressure}
\end{figure}

Figure~\ref{fig:001_Pb54_on_Ar_PS_Lifetime_over_avg_Pressure} displays the fitted Pb$^{54+}$ ion beam lifetimes $\tau$ from Eq.~\eqref{eq:beam_intensity} as a function of the estimated average target pressure $\bar{P}$. Also shown is the predicted lifetime from \verb|beam_gas_collisions| of Pb$^{54+}$ on Ar, assuming that the pressure bump from the injected neutral gas dominates. Below $10^{-9}$ mbar, the injected target gas composition is considered too uncertain as total flows at these Set Points are not known and outgassing is not identical for all experiments. Increasing the average PS pressure provokes a clearly decreasing trend in measured lifetime. There is no apparent dependence on either longitudinal beam profile (bunched/coasting), or on beam intensity. The horizontal error bars in the estimated target pressure represent the saturated/unsaturated bounds discussed in Sec.~\ref{sec:pressure_profile_reconstruction}, although the pressure errors may be much larger. The vertical error bars in the estimated lifetime are based on shot-to-shot variations for each Set Point and the lifetime parameter fitting error and are much smaller relative to the pressure uncertainty. 

The semi-empirical model prediction for Pb$^{54+}$ on Ar approaches the measurements as the average target pressure increases. A grey interval illustrates the (at least) factor 2 uncertainty mentioned by Shevelko~\cite{shevelko2011_summary} in the underlying models for electron capture and loss. As more target gas is injected, beam-gas interaction becomes the primary loss mechanism and the measured lifetimes converge with the predictions. An alternative is that the underlying semi-empirical models are less accurate for lower pressures. 

\subsection{Mg\texorpdfstring{$^{7+}$}{Lg} lifetime experiment with injected Ar gas}
\label{sec:Mg7_on_ar}

\begin{figure}[t]
\centering
\includegraphics[width=\columnwidth]{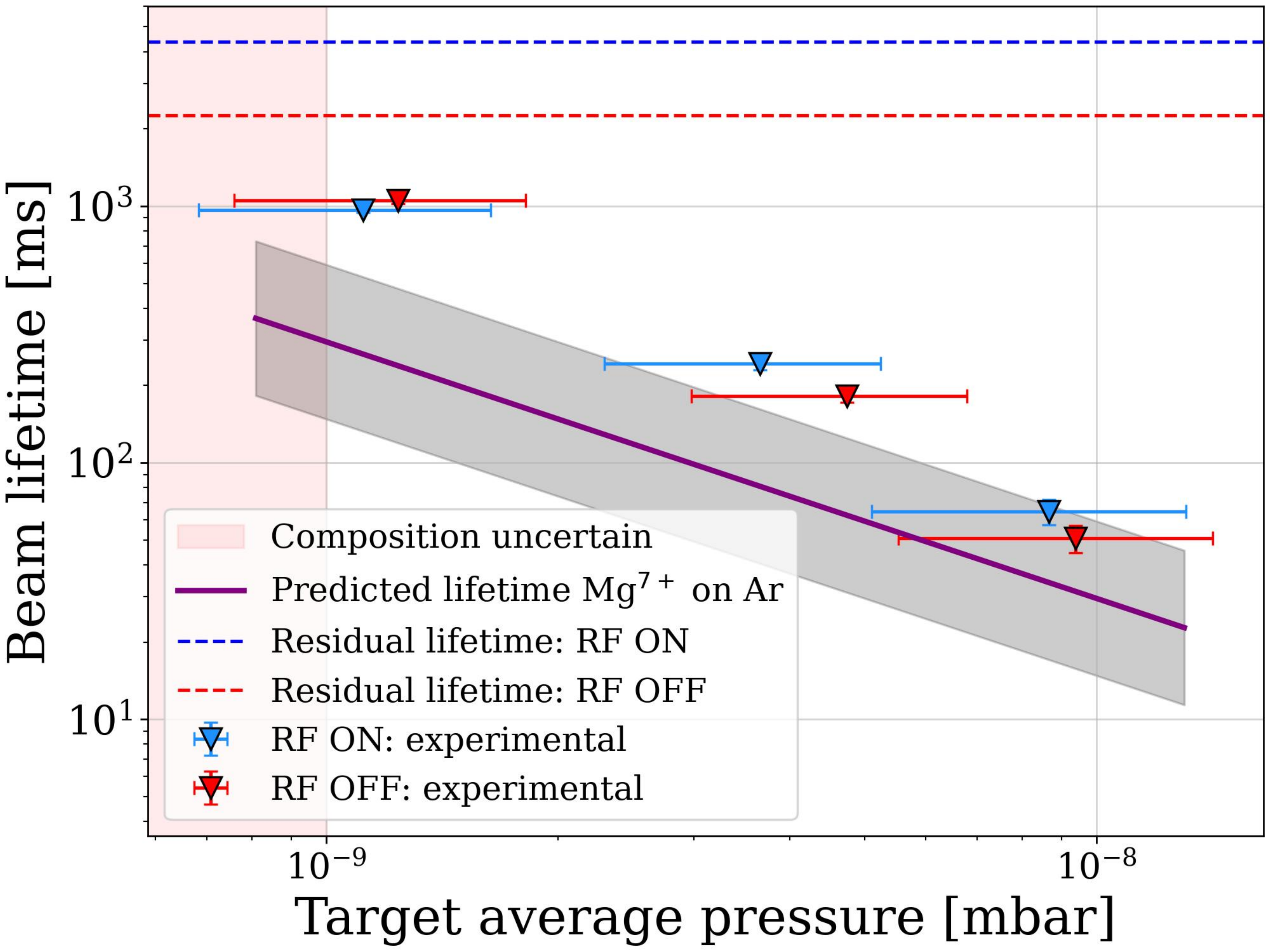}
\caption{Calculated and measured Mg$^{7+}$ beam lifetimes for various quantities of injected Ar gas. The residual lifetimes without any gas injection are also shown.}
\label{fig:002_Mg7_on_Ar_PS_Lifetime_over_avg_Pressure}
\end{figure}

The second PS lifetime experiment (B) was conducted in May 2024 with a test beam of Mg$^{7+}$ at 90.2 MeV/u, with argon as the gas target. This beam was produced in the source and LINAC3 for the very first time. Although hardly comparable to the operational routine Pb$^{54+}$ beams, the lifetime experiments made use of the available Mg beam intensities of the EARLY type, following a similar procedure as with Pb$^{54+}$ on Ar described in Sec~\ref{sec:Pb54_on_Ar}. High noise levels in the BCT readout for this low-intensity beam allowed us to observe a clear impact from neutral Ar gas injection only for the three highest Set Points, shown in Fig.~\ref{fig:002_Mg7_on_Ar_PS_Lifetime_over_avg_Pressure}. Nonetheless, a decreasing lifetime trend appears. Conversely to the Pb$^{54+}$ experiment (A), the measurements approach the predicted Mg$^{7+}$ lifetime on Ar from above. For the highest pressures, the measured lifetime falls within the prediction limits.

In this scenario, the combination of a loosely bound outer projectile electron, high $Z_T$ (meaning higher than H, He, and N), and relativistic velocity is not represented in Fig.~\ref{fig:EL_new_combined_semi_empirical_formula}. Consequently, the empirical parameter optimization for Table~\ref{tab:fitting_coefficients_semiempirical} in~\cite{weber_2016_semi_empirical_formula} to match experimental data was probably done without reference points in this experimental regime. In addition, EC cross sections for high $Z$ targets such as Ar could be overestimated, although EL is the dominant mechanism in this regime. Nonetheless, the lifetime predictions still fall within a factor two of the measurement for the highest pressure.

\subsection{Pb\texorpdfstring{$^{54+}$}{Lg} lifetime experiment with injected He gas}
\label{sec:Pb54_on_He}

\begin{figure}[t]
\centering
\includegraphics[width=\columnwidth]{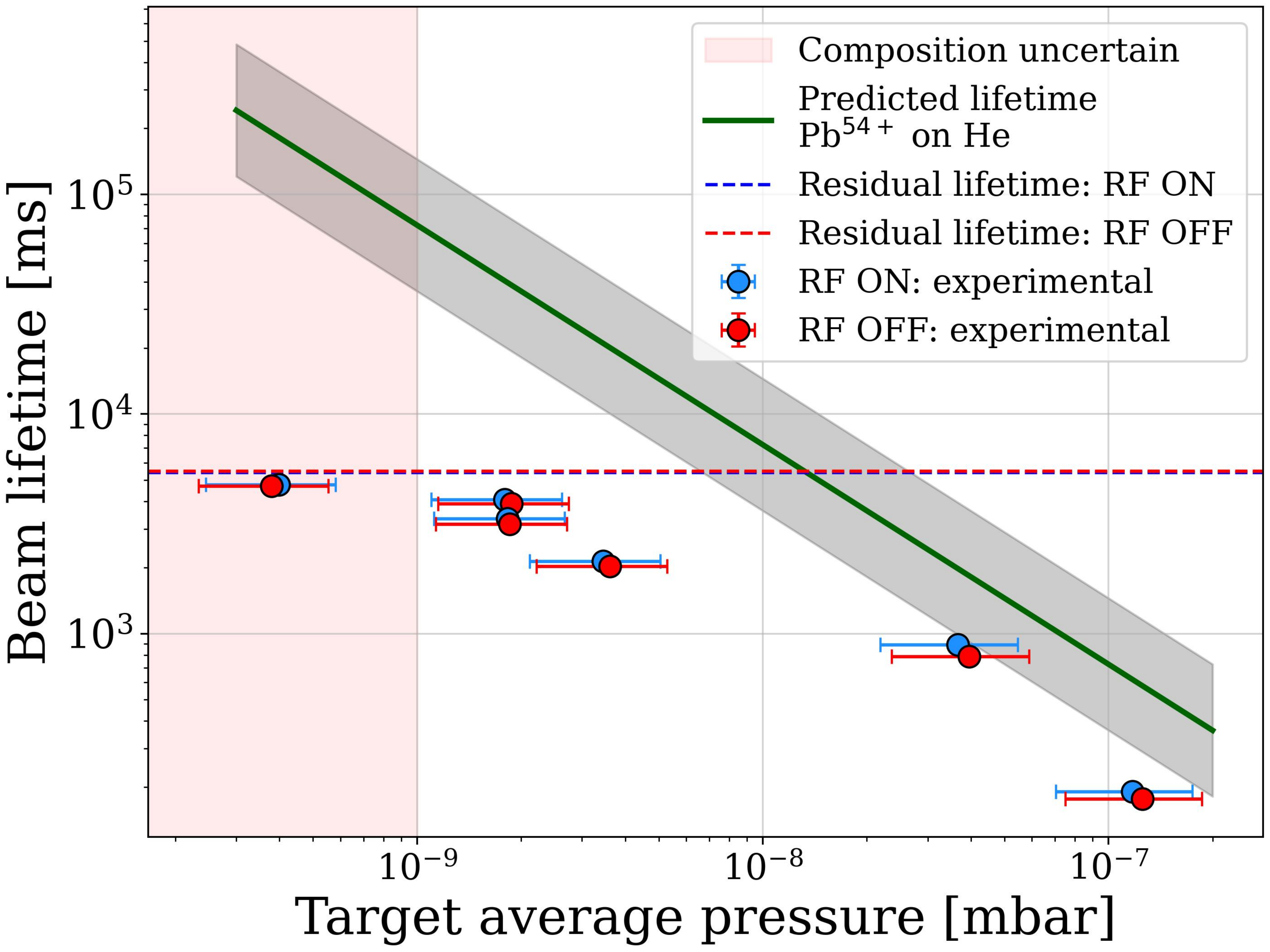}
\caption{Calculated and measured Pb$^{54+}$ beam lifetimes for different quantities of injected He gas, also showing the residual lifetimes without any gas injection.}
\label{fig:003_Pb54_on_He_PS_Lifetime_over_avg_Pressure}
\end{figure}

\begin{figure*}[t]
    \centering    \includegraphics[width=\textwidth]{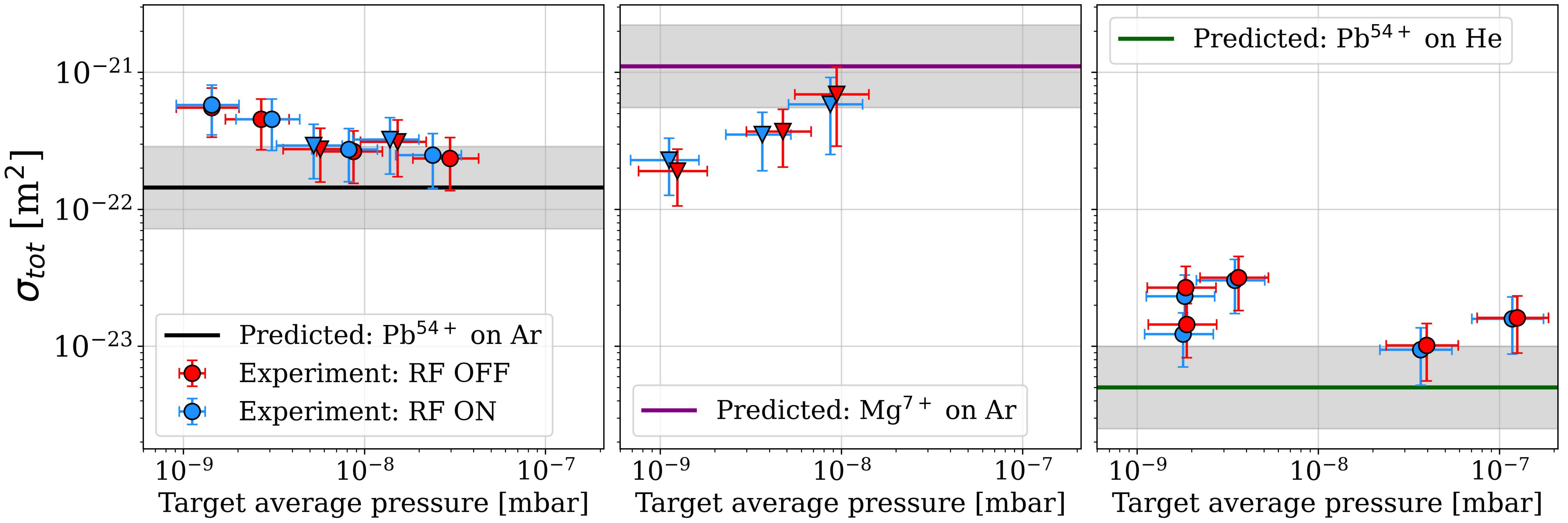}
    \caption{Experimental and calculated (predicted from the semi-empirical model) total cross sections for the three PS experiments: (A) Pb$^{54+}$ at 72.2 MeV/u on Ar, (B) Mg$^{7+}$ at 90.2 MeV/u on Ar, and (C) Pb$^{54+}$ at 72.2 MeV/u on He.} %  The PS baseline pressure is highlighted, below which the contribution to the PS average pressure is negligible. Most likely, the limit above which the beam losses from the pressure bump dominates with respect to the background is $5 \times 10^{-9}$ mbar.
    \label{fig:All_PS_Sigma_over_avg_Pressure_adjusted_residual_pressure}
\end{figure*}

The third experiment (C) in this study features Pb$^{54+}$ beams at 72.2 MeV/u with injected neutral He gas, carried out in September 2024. The resulting measured beam lifetimes as a function of pressure are shown in Fig.~\ref{fig:003_Pb54_on_He_PS_Lifetime_over_avg_Pressure}. The predicted lifetime of Pb$^{54+}$ on He is about 1.9 times higher than the measured lifetime for the second-highest SP. Compared to the high $Z_T$ Ar target in experiments (A) and (B), EC and EL cross sections for Pb$^{54+}$ on He are several orders of magnitude smaller. Hence, larger amounts of injected He gas are required compared to Ar before any clear impact on beam lifetime is noticed. 

\subsection{Experimental cross sections from beam lifetimes}

\begin{table}[b]
\begin{ruledtabular}
\begin{tabular}{l|lll}
Experiment & A & B & C \\ \hline
Date & 10/2023 & 05/2024 & 09/2024 \\
Projectile & Pb$^{54+}$ & Mg$^{7+}$ & Pb$^{54+}$ \\ 
$E_k$ [MeV/u] & 72.2 & 90.2 & 72.2 \\
Target  & Ar & Ar & He \\
RF ON: ${\tau_{\textrm{res}}}$ [s] & 5.9 $\pm$ 0.2 & 4.4 $\pm$ 1.1 & 5.4 $\pm$ 0.1 \\
RF OFF: ${\tau_{\textrm{res}}}$ [s]  & 6.1 $\pm$ 0.1 & 2.3 $\pm$ 0.7 & 5.5 $\pm$ 0.1 \\ \hline
$\sigma_{\text{pred.}}$ [$\times10^{-22}$ m$^2$] & $1.4 \pm 0.7$ & $11 \pm 6$ & $0.05 \pm 0.03$ \\
$\sigma_{\text{exp. }}$ [$\times10^{-22}$ m$^2$] & $3.6 \pm 1.5$ & $4.0 \pm 2.1$  & $0.19 \pm 0.08$ \\
\end{tabular}
\caption{Experimental and predicted (calculated) cross sections from the semi-empirical prediction model.}
\label{table:projectiles_and_energies}
\end{ruledtabular}
\end{table}

Next, we compute experimental cross sections $\sigma_{\text{exp.}}$ from measured beam-gas interactions lifetimes, also comparing with the calculated semi-empirical model predictions $\sigma_{\text{pred.}}$ from \verb|beam_gas_collisions|. Measured beam lifetimes and average pressures are used in Eq.~\eqref{eq:beam_lifetime} inversely. The baseline decay constant (from the residual lifetimes) was subtracted according to Eq.~\eqref{eq:tau_bump_lifetime_addition_2}. Only lifetimes with target pressure above $10^{-9}$ mbar (where the gas target composition is better known) are included in the calculations. The prediction model cross section error is calculated from the estimated lifetime and pressure errors using the generalized error propagation formula in Eq.~(2.7) from~\cite{error_propagation_formula}. The cross section results are shown in Fig.~\ref{fig:All_PS_Sigma_over_avg_Pressure_adjusted_residual_pressure}.

Table~\ref{table:projectiles_and_energies} summarizes the gas injection experiments with projectile types, projectile kinetic energy $E_k$, target gas type, calculated total cross section from the semi-empirical prediction model, ${\tau_{\textrm{residual}}}$ from Eq.~\eqref{eq:tau_bump_lifetime_addition_2}, both for RF ON and RF OFF, and finally cross sections $\sigma_{\text{pred.}}$, and $\sigma_{\text{exp.}}$. In general, we observe a convergence of $\sigma_{\text{exp.}}$ towards $\sigma_{\text{pred.}}$ for higher pressures, where losses from beam-gas interactions start to dominate compared to other sources; the trend is similar to Fig.~\ref{fig:001_Pb54_on_Ar_PS_Lifetime_over_avg_Pressure}, \ref{fig:002_Mg7_on_Ar_PS_Lifetime_over_avg_Pressure} and \ref{fig:003_Pb54_on_He_PS_Lifetime_over_avg_Pressure}. The fact that predicted Pb$^{54+}$ cross sections typically fall below the measurements, conversely to the Mg$^{7+}$ on Ar prediction which falls above, is simply a manifestation of these lifetime trends but propagated to the cross sections.

Also, the longitudinal beam profile --- bunched (RF ON) or coasting (RF OFF) --- seems to have no impact on beam lifetime or cross section in these experiments, confirming that charge-changing beam-gas interactions become the main source of intensity loss with higher injected target gas amounts. For higher target pressures, $\sigma_{\text{exp.}}$ falls into the uncertainty band of the semi-empirical $\sigma_{\text{pred.}}$, well within measured lifetimes by a factor 2-3 or less, the typical value reported in Ref.~\cite{shevelko2011_summary}, when beam-gas interactions become the dominant loss mechanism. 

Regarding electron capture, it is worth mentioning that only single-electron capture processes are considered in the Schlachter formula in Eq.~\eqref{eq:schlachter} ---implemented in \verb|beam_gas_collisions| --- not multi-electron loss processes. These physical processes require more complex simulation codes such as \texttt{CAPTURE}, \texttt{DEPOSIT} or \texttt{RICODE-M}, but are rare in comparison to single-electron capture and are mostly important up to only a few MeV/u. Multi-electron processes should be investigated further for LEIR, with present injection energies of 4.2 MeV/u for Pb ions, but should matter less for the PS and the SPS.

A full overview of all predicted EC and EL cross sections for future ion species on common target gases in LEIR and the PS is found in the Appendix, calculated using \verb|beam_gas_collisions|. In the context of future ion species and beam-gas interactions, we encourage further experimental lifetime and vacuum studies.

\section{Summary and Conclusions}
\label{sec:conclusions} 

In this study, we present the publicly accessible scientific Python package \verb|beam_gas_collisions| to calculate (1) atomic cross sections, and (2) beam lifetimes for projectiles interacting with residual gas in a generic accelerator. Two main effects are considered: electron capture and electron loss. The electron loss model is based on new numerical implementations that refine the fitting parameters from previous semi-empirical studies by Weber, Shevelko, and DuBois. These studies cover a broad range of target atomic numbers $Z_T$ , projectile types, and energies. The electron capture model uses the established Schlachter semi-empirical formula. 

The beam lifetime model was benchmarked in three neutral gas injection experiments in the PS: (A) Pb$^{54+}$ on Ar at 72.2 MeV/u, (B) Mg$^{7+}$ on Ar at 90.2 MeV/u and (C) Pb$^{54+}$ on He at 72.2 MeV/u. A controlled pressure bump of injected gas around the BGI instrumentation was gradually increased to generate an in-ring gas target, and its impact on the ion beam lifetime was measured. A clear beam lifetime decrease was observed for higher target pressures, seemingly independent of beam intensity, tunes, and longitudinal profiles. A 3D-geometry simulation effort of the PS beamline around the BGI instrumentation was carried out to estimate the pressure bump profile and the average PS pressure. Due to high uncertainties in the baseline residual gas, the decay constant --- from beam lifetime measurements before inserting the controlled pressure bump --- was factored out. For higher pressures, beam-gas interactions become the dominant loss mechanism, and the measured lifetime trend converges toward the predicted value. At lower pressures, full knowledge of other background loss sources is not yet clear. For every such experiment, we also report the experimental cross section values as a function of pressure. For the higher pressure values, the predicted lifetime falls within the factor 2-3 prediction model uncertainty reported by previous studies. 

The largest source of uncertainty remains the PS average pressure estimate, as the pressure and composition measurements are local and can have large variations along the 628 m of the PS ring. To reduce the uncertainties, a considerable effort to characterize several positions along the beamline and a better model is required but is out of the scope of the present study. Hence, the average PS pressure error bars are likely larger than the reported $\pm$30\%  accuracy of the gauges used in the machine. Other factors of uncertainty unique to the Mg$^{7+}$ on Ar experiment mainly include the low beam intensity, limited data points, and a prediction model under-representation of scenarios with loosely bound outer electron and high $Z_T$ for the fitting parameters. These elements could be addressed in future studies. Nonetheless, the overall prediction-measurement agreement is notable considering that the PS experimental set-up around the BGI in these measurement campaigns is not a dedicated gas target. Improved future studies will most likely require such a dedicated setup and additional hardware. We also encourage further benchmarking experiments of this combined semi-empirical prediction model with additional projectiles, target gases, and energy ranges, as well as in other experimental setups and accelerators.

\begin{acknowledgments}
We acknowledge the support of LINAC3, LEIR and PS operation teams as well as J. Storey during the target gas injection experiment. We gratefully acknowledge fruitful discussions with  C. Carli, H. Bartosik, D. Gamba, W. Krasny, D. K\"{u}chler, E. Mahner,  R. Scrivens and M. Seidel on these topics. These experiments and studies were made possible through support of the CERN BE-ABP group, which the authors acknowledge.
\end{acknowledgments}

\section*{Author contribution}
E.W. coordinated the \verb|beam_gas_collisions| package development, experimental post-processing, and the manuscript writing. E.W., J.O., R.A.F., and J.S. conducted experiment (A) and the related analysis. E.W., F.U., R.A.F., and J.S. conducted experiments (B) and (C). F.U. and J.S. developed the final vacuum and pressure profile modelling. G.W. provided vital input for the semi-empirical EL formula theory and historical benchmarking data. All authors actively contributed to improving the manuscript during several iterations. % E.W. performed the calculations on future ion species.

\newpage

\appendix
% Numbering A.x
\newcommand{\hbAppendixPrefix}{A}
\renewcommand{\thefigure}{\hbAppendixPrefix.\arabic{figure}}
\setcounter{figure}{0}
\renewcommand{\thetable}{\hbAppendixPrefix.\arabic{table}} 
\setcounter{table}{0}
\renewcommand{\theequation}{\hbAppendixPrefix.\arabic{equation}} 
\setcounter{equation}{0}

\onecolumngrid
\section{Cross section calculations of electron capture and loss}
\label{sec:appendix}

\begin{figure*}[h]
    \centering
    \includegraphics[width=\textwidth]{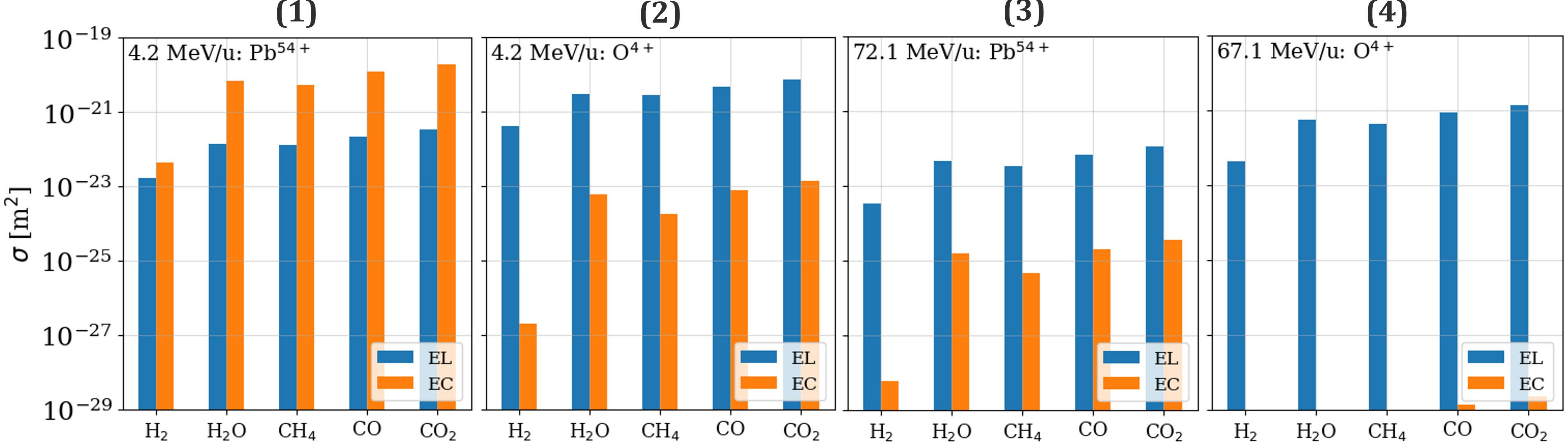}
    \caption{Calculated EC and EL cross sections on H$_2$, H$_2$O, CH$_4$, CO and CO$_2$ gas targets for the projectiles: 1) Pb$^{54+}$ at 4.2 MeV/u, 2) O$^{4+}$ at 4.2 MeV/u, 3) Pb$^{54+}$ at 72.1 MeV/u and 4) O$^{4+}$ at 67.1 MeV/u. The temperature is set to $T = 293$ K.}
    \label{fig:leir_ps_cross_sections}
\end{figure*}

Table~\ref{table:cross_sections_leir_ps} presents the calculated EC cross sections $\sigma_{\text{EC}}$ from Eq.~\eqref{eq:schlachter} and EL cross sections $\sigma_{\text{EL}}$ from Eq.~\eqref{eq:new_EL_semiempirical} for energies and projectiles relevant in LEIR and the PS, using \verb|beam_gas_collisions|. Figure~\ref{fig:leir_ps_cross_sections} graphically illustrates cross sections for the study cases Pb$^{54+}$ and O$^{4+}$ on the common gas target types H$_2$, H$_2$O, CH$_4$, CO and CO$_2$ . In LEIR, EC processes are more important for Pb$^{54+}$ and EL processes are more important for O$^{4+}$. In the PS, note the high EL cross sections for O$^{4+}$, compared to Pb$^{54+}$.

\begin{table}[h]
\begin{ruledtabular}
\begin{tabular}{l|l|l|l|l|l|l||l|l|l|l|l|l}
           &                & \multicolumn{5}{c||}{EC cross section {[}m$^2${]}} & \multicolumn{5}{c}{EL cross section {[}m$^2${]}} \\ \hline
Projectile & $E_k$ [MeV/u] & H$_2$    & H$_2$O  & CH$_4$  & CO      & CO$_2$  & H$_2$    & H$_2$O  & CH$_4$  & CO      & CO$_2$  \\ \hline
He$^{1+}$  & 4.2            & 3.3E-29  & 1.0E-25 & 3.0E-26 & 1.3E-25 & 2.3E-25 & 3.3E-22  & 2.3E-21 & 2.2E-21 & 3.6E-21 & 5.6E-21 \\
He$^{2+}$  & 4.2            & 2.7E-28  & 8.1E-25 & 2.4E-25 & 1.1E-24 & 1.9E-24 & -        & -       & -       & -       & -       \\
O$^{4+}$   & 4.2            & 2.0E-27  & 5.8E-24 & 1.7E-24 & 7.6E-24 & 1.3E-23 & 4.0E-22  & 2.9E-21 & 2.8E-21 & 4.5E-21 & 7.0E-21 \\
O$^{5+}$   & 4.2            & 4.8E-27  & 1.4E-23 & 4.1E-24 & 1.8E-23 & 3.1E-23 & 4.1E-22  & 3.0E-21 & 2.9E-21 & 4.6E-21 & 7.2E-21 \\
O$^{8+}$   & 4.2            & 2.7E-26  & 6.9E-23 & 2.5E-23 & 9.4E-23 & 1.6E-22 & -        & -       & -       & -       & -       \\
Mg$^{6+}$  & 4.2            & 9.7E-27  & 2.6E-23 & 8.4E-24 & 3.5E-23 & 6.1E-23 & 3.1E-22  & 2.3E-21 & 2.2E-21 & 3.5E-21 & 5.5E-21 \\
Mg$^{7+}$  & 4.2            & 1.8E-26  & 4.5E-23 & 1.5E-23 & 6.0E-23 & 1.0E-22 & 2.1E-22  & 1.6E-21 & 1.5E-21 & 2.4E-21 & 3.8E-21 \\
Ar$^{11+}$ & 4.2            & 9.3E-26  & 1.8E-22 & 7.9E-23 & 2.6E-22 & 4.3E-22 & 1.2E-22  & 9.0E-22 & 8.4E-22 & 1.4E-21 & 2.2E-21 \\
Ca$^{17+}$ & 4.2            & 5.0E-25  & 5.4E-22 & 3.1E-22 & 8.5E-22 & 1.4E-21 & 2.9E-23  & 2.3E-22 & 2.2E-22 & 3.6E-22 & 5.7E-22 \\
Kr$^{22+}$ & 4.2            & 1.4E-24  & 9.9E-22 & 6.3E-22 & 1.6E-21 & 2.6E-21 & 8.2E-23  & 6.6E-22 & 6.1E-22 & 1.0E-21 & 1.6E-21 \\
In$^{37+}$ & 4.2            & 1.0E-23  & 3.1E-21 & 2.2E-21 & 5.3E-21 & 8.4E-21 & 1.1E-23  & 9.5E-23 & 8.8E-23 & 1.5E-22 & 2.3E-22 \\
Xe$^{39+}$ & 4.2            & 1.2E-23  & 3.5E-21 & 2.5E-21 & 5.9E-21 & 9.4E-21 & 1.2E-23  & 1.0E-22 & 9.2E-23 & 1.6E-22 & 2.5E-22 \\
Pb$^{54+}$ & 4.2            & 4.3E-23  & 6.9E-21 & 5.1E-21 & 1.2E-20 & 1.9E-20 & 1.6E-23  & 1.4E-22 & 1.3E-22 & 2.2E-22 & 3.5E-22 \\ \hline \hline
He$^{1+}$  & 67             & 5.6e-35  & 1.7e-31 & 5.2e-32 & 2.2e-31 & 4.0e-31 & 3.7E-23  & 4.6E-22 & 3.6E-22 & 7.1E-22 & 1.1E-21 \\
He$^{2+}$  & 245.4          & 9.0e-37  & 2.8e-33 & 8.4e-34 & 3.6e-33 & 6.4e-33 & -        & -       & -       & -       & -       \\
O$^{4+}$   & 67.1           & 3.3e-33  & 1.0E-29 & 3.1e-30 & 1.3E-29 & 2.4E-29 & 4.4E-23  & 5.6E-22 & 4.3E-22 & 8.5E-22 & 1.4E-21 \\
O$^{5+}$   & 102.9          & 1.0e-33  & 3.2e-30 & 9.5e-31 & 4.1e-30 & 7.3e-30 & 3.7E-23  & 5.9E-22 & 4.1E-22 & 9.0E-22 & 1.5E-21 \\
O$^{8+}$   & 245.6          & 9.8e-35  & 3.0e-31 & 9.1e-32 & 4.0e-31 & 7.0e-31 & -        & -       & -       & -       & -       \\
Mg$^{6+}$  & 67.1           & 1.6e-32  & 5.0E-29 & 1.5E-29 & 6.5E-29 & 1.2E-28 & 3.6E-23  & 4.5E-22 & 3.5E-22 & 6.9E-22 & 1.1E-21 \\
Mg$^{7+}$  & 90.2           & 7.1e-33  & 2.2E-29 & 6.6e-30 & 2.9E-29 & 5.1E-29 & 2.0E-23  & 3.2E-22 & 2.3E-22 & 4.9E-22 & 7.9E-22 \\
Ar$^{11+}$ & 80.7           & 7.1e-32  & 2.2E-28 & 6.6E-29 & 2.9E-28 & 5.1E-28 & 1.4E-23  & 1.9E-22 & 1.4E-22 & 3.0E-22 & 4.8E-22 \\
Ca$^{17+}$ & 183            & 7.6e-33  & 2.4E-29 & 7.1e-30 & 3.1E-29 & 5.4E-29 & 3.1E-24  & 5.3E-23 & 3.6E-23 & 7.9E-23 & 1.3E-22 \\
Kr$^{22+}$ & 70.2           & 2.1e-30  & 5.7E-27 & 1.7E-27 & 7.4E-27 & 1.3E-26 & 1.1E-23  & 1.5E-22 & 1.1E-22 & 2.2E-22 & 3.6E-22 \\
In$^{37+}$ & 108.8          & 1.9e-30  & 5.2E-27 & 1.5E-27 & 6.7E-27 & 1.2E-26 & 2.3E-24  & 3.8E-23 & 2.7E-23 & 5.8E-23 & 9.4E-23 \\
Xe$^{39+}$ & 96.7           & 4.2e-30  & 1.1E-26 & 3.3E-27 & 1.4E-26 & 2.6E-26 & 2.9E-24  & 4.7E-23 & 3.3E-23 & 7.1E-23 & 1.2E-22 \\
Pb$^{54+}$ & 72.1           & 6.0E-29  & 1.6E-25 & 4.8E-26 & 2.1E-25 & 3.7E-25 & 3.4E-24  & 4.6E-23 & 3.5E-23 & 7.0E-23 & 1.1E-22
\end{tabular}
\end{ruledtabular}
\caption{Calculated $\sigma_{\text{EC}}$ and $\sigma_{\text{EL}}$ on target gases H$_2$, H$_2$O, CH$_4$, CO and CO$_2$ for projectile energies relevant in LEIR (top half of table) and in the PS (bottom half of table).}
\label{table:cross_sections_leir_ps}
\end{table}

\twocolumngrid
\newpage
%\clearpage 
\bibliography{references}% Produces the bibliography via BibTeX.

\begin{thebibliography}{58}%
\makeatletter
\providecommand \@ifxundefined [1]{%
 \@ifx{#1\undefined}
}%
\providecommand \@ifnum [1]{%
 \ifnum #1\expandafter \@firstoftwo
 \else \expandafter \@secondoftwo
 \fi
}%
\providecommand \@ifx [1]{%
 \ifx #1\expandafter \@firstoftwo
 \else \expandafter \@secondoftwo
 \fi
}%
\providecommand \natexlab [1]{#1}%
\providecommand \enquote  [1]{``#1''}%
\providecommand \bibnamefont  [1]{#1}%
\providecommand \bibfnamefont [1]{#1}%
\providecommand \citenamefont [1]{#1}%
\providecommand \href@noop [0]{\@secondoftwo}%
\providecommand \href [0]{\begingroup \@sanitize@url \@href}%
\providecommand \@href[1]{\@@startlink{#1}\@@href}%
\providecommand \@@href[1]{\endgroup#1\@@endlink}%
\providecommand \@sanitize@url [0]{\catcode `\\12\catcode `\$12\catcode `\&12\catcode `\#12\catcode `\^12\catcode `\_12\catcode `\%12\relax}%
\providecommand \@@startlink[1]{}%
\providecommand \@@endlink[0]{}%
\providecommand \url  [0]{\begingroup\@sanitize@url \@url }%
\providecommand \@url [1]{\endgroup\@href {#1}{\urlprefix }}%
\providecommand \urlprefix  [0]{URL }%
\providecommand \Eprint [0]{\href }%
\providecommand \doibase [0]{https://doi.org/}%
\providecommand \selectlanguage [0]{\@gobble}%
\providecommand \bibinfo  [0]{\@secondoftwo}%
\providecommand \bibfield  [0]{\@secondoftwo}%
\providecommand \translation [1]{[#1]}%
\providecommand \BibitemOpen [0]{}%
\providecommand \bibitemStop [0]{}%
\providecommand \bibitemNoStop [0]{.\EOS\space}%
\providecommand \EOS [0]{\spacefactor3000\relax}%
\providecommand \BibitemShut  [1]{\csname bibitem#1\endcsname}%
\let\auto@bib@innerbib\@empty
%</preamble>
\bibitem [{\citenamefont {Benedikt}\ \emph {et~al.}(2004)\citenamefont {Benedikt}, \citenamefont {Collier}, \citenamefont {Mertens}, \citenamefont {Poole},\ and\ \citenamefont {Schindl}}]{benedikt_lhc_2004}%
  \BibitemOpen
  \bibfield  {author} {\bibinfo {author} {\bibfnamefont {M.}~\bibnamefont {Benedikt}}, \bibinfo {author} {\bibfnamefont {P.}~\bibnamefont {Collier}}, \bibinfo {author} {\bibfnamefont {V.}~\bibnamefont {Mertens}}, \bibinfo {author} {\bibfnamefont {J.}~\bibnamefont {Poole}},\ and\ \bibinfo {author} {\bibfnamefont {K.}~\bibnamefont {Schindl}},\ }\href {https://doi.org/10.5170/CERN-2004-003-V-3} {{\selectlanguage {sv}\bibinfo {title} {{LHC} {Design} {Report}}}} (\bibinfo {year} {2004}),\ \bibinfo {note} {iSBN: 9789290832393 Number: CERN-2004-003-V-3 Publisher: CERN}\BibitemShut {NoStop}%
\bibitem [{\citenamefont {{J. Coupard}}\ \emph {et~al.}(2016)\citenamefont {{J. Coupard}}, \citenamefont {Damerau}, \citenamefont {Funken}, \citenamefont {Garoby} \emph {et~al.}}]{j_coupard_lhc_2016}%
  \BibitemOpen
  \bibfield  {author} {\bibinfo {author} {\bibnamefont {{J. Coupard}}}, \bibinfo {author} {\bibfnamefont {H.}~\bibnamefont {Damerau}}, \bibinfo {author} {\bibfnamefont {A.}~\bibnamefont {Funken}}, \bibinfo {author} {\bibfnamefont {R.}~\bibnamefont {Garoby}}, \emph {et~al.},\ }\href {https://edms.cern.ch/ui/file/1626950/1/LIU-ION-TDR_v8_docx_cpdf.pdf} {\emph {\bibinfo {title} {{LHC} {Injectors} {Upgrade}: {Technical} {Design} {Report} {Vol}. {II}: {Ions}}}},\ \bibinfo {type} {Technical {Report}}\ \bibinfo {number} {EDMS 1626950}\ (\bibinfo {year} {2016})\BibitemShut {NoStop}%
\bibitem [{\citenamefont {Lopienska}(2022)}]{lopienska2022_cern_complex}%
  \BibitemOpen
  \bibfield  {author} {\bibinfo {author} {\bibfnamefont {E.}~\bibnamefont {Lopienska}},\ }\href@noop {} {} (\bibinfo {year} {2022}),\ \bibinfo {note} {the CERN accelerator complex (2022 layout), available at: \url{https://cds.cern.ch/record/2800984/files/CCC-v2022.png}}\BibitemShut {NoStop}%
\bibitem [{\citenamefont {Argyropoulos}\ \emph {et~al.}(2019)\citenamefont {Argyropoulos} \emph {et~al.}}]{argyropoulos_slip_stacking_2019}%
  \BibitemOpen
  \bibfield  {author} {\bibinfo {author} {\bibfnamefont {T.}~\bibnamefont {Argyropoulos}} \emph {et~al.},\ }\bibfield  {title} {{\selectlanguage {en}\bibinfo {title} {Momentum {Slip}-{Stacking} in {CERN} {SPS} for the {Ion} {Beams}}},\ }\href {https://doi.org/10.18429/JACOW-IPAC2019-WEPTS039} {\bibfield  {journal} {\bibinfo  {journal} {Proceedings of the 10th Int. Particle Accelerator Conf.}\ }\textbf {\bibinfo {volume} {IPAC2019}},\ \bibinfo {pages} {4 pages, 0.777 MB} (\bibinfo {year} {2019})},\ \bibinfo {note} {artwork Size: 4 pages, 0.777 MB ISBN: 9783954502080 Medium: PDF Publisher: JACoW Publishing, Geneva, Switzerland}\BibitemShut {NoStop}%
\bibitem [{\citenamefont {Bartosik}\ \emph {et~al.}(2021)\citenamefont {Bartosik} \emph {et~al.}}]{bartosik_injectors_2021}%
  \BibitemOpen
  \bibfield  {author} {\bibinfo {author} {\bibfnamefont {H.}~\bibnamefont {Bartosik}} \emph {et~al.},\ }\bibfield  {title} {{\selectlanguage {en}\bibinfo {title} {Injectors beam performance evolution during run 2}},\ }in\ \href {https://cds.cern.ch/record/2750276} {{\selectlanguage {en}\emph {\bibinfo {booktitle} {Proceedings of The 2019 Evian Workshop On {LHC} Beam Operations}}}}\ (\bibinfo {year} {2021})\BibitemShut {NoStop}%
\bibitem [{\citenamefont {Bartosik}\ \emph {et~al.}(2017)\citenamefont {Bartosik} \emph {et~al.}}]{bartosik_lhc_2017}%
  \BibitemOpen
  \bibfield  {author} {\bibinfo {author} {\bibfnamefont {H.}~\bibnamefont {Bartosik}} \emph {et~al.},\ }\bibfield  {title} {{\selectlanguage {en}\bibinfo {title} {The {LHC} {Injectors} {Upgrade} ({LIU}) {Project} at {CERN}: {Ion} {Injector} {Chain}}},\ }\href@noop {} {\bibfield  {journal} {\bibinfo  {journal} {Proceedings of the 8th Int. Particle Accelerator Conf.}\ }\textbf {\bibinfo {volume} {IPAC2017}},\ \bibinfo {pages} {4} (\bibinfo {year} {2017})},\ \bibinfo {note} {iSBN: 978-3-95450-182-3}\BibitemShut {NoStop}%
\bibitem [{\citenamefont {Franchetti}\ \emph {et~al.}(2010)\citenamefont {Franchetti}, \citenamefont {Chorniy}, \citenamefont {Hofmann}, \citenamefont {Bayer}, \citenamefont {Becker}, \citenamefont {Forck}, \citenamefont {Giacomini}, \citenamefont {Kirk}, \citenamefont {Mohite}, \citenamefont {Omet}, \citenamefont {Parfenova},\ and\ \citenamefont {Schütt}}]{franchetti_experiment_2010}%
  \BibitemOpen
  \bibfield  {author} {\bibinfo {author} {\bibfnamefont {G.}~\bibnamefont {Franchetti}}, \bibinfo {author} {\bibfnamefont {O.}~\bibnamefont {Chorniy}}, \bibinfo {author} {\bibfnamefont {I.}~\bibnamefont {Hofmann}}, \bibinfo {author} {\bibfnamefont {W.}~\bibnamefont {Bayer}}, \bibinfo {author} {\bibfnamefont {F.}~\bibnamefont {Becker}}, \bibinfo {author} {\bibfnamefont {P.}~\bibnamefont {Forck}}, \bibinfo {author} {\bibfnamefont {T.}~\bibnamefont {Giacomini}}, \bibinfo {author} {\bibfnamefont {M.}~\bibnamefont {Kirk}}, \bibinfo {author} {\bibfnamefont {T.}~\bibnamefont {Mohite}}, \bibinfo {author} {\bibfnamefont {C.}~\bibnamefont {Omet}}, \bibinfo {author} {\bibfnamefont {A.}~\bibnamefont {Parfenova}},\ and\ \bibinfo {author} {\bibfnamefont {P.}~\bibnamefont {Schütt}},\ }\bibfield  {title} {\bibinfo {title} {Experiment on space charge driven nonlinear resonance crossing in an ion synchrotron},\ }\href {https://doi.org/10.1103/PhysRevSTAB.13.114203} {\bibfield  {journal} {\bibinfo  {journal} {Physical Review
  Special Topics - Accelerators and Beams}\ }\textbf {\bibinfo {volume} {13}},\ \bibinfo {pages} {114203} (\bibinfo {year} {2010})},\ \bibinfo {note} {publisher: American Physical Society}\BibitemShut {NoStop}%
\bibitem [{\citenamefont {Antoniou Et~Al.}(2017)}]{antoniou_et_al_transverse_2017}%
  \BibitemOpen
  \bibfield  {author} {\bibinfo {author} {\bibfnamefont {F.}~\bibnamefont {Antoniou Et~Al.}},\ }\bibfield  {title} {{\selectlanguage {en}\bibinfo {title} {Transverse studies with ions at {SPS} flat bottom}},\ }\href {https://doi.org/10.23727/CERN-PROCEEDINGS-2017-002.145} {\bibfield  {journal} {\bibinfo  {journal} {CERN Proceedings}\ ,\ \bibinfo {pages} {145 Pages}} (\bibinfo {year} {2017})},\ \bibinfo {note} {artwork Size: 145 Pages Publisher: CERN Proceedings}\BibitemShut {NoStop}%
\bibitem [{\citenamefont {Hernandez}\ \emph {et~al.}(2018)\citenamefont {Hernandez}, \citenamefont {Moreno}, \citenamefont {Bartosik}, \citenamefont {Biancacci}, \citenamefont {Hirlander},\ and\ \citenamefont {Huschauer}}]{hernandez_space_2018}%
  \BibitemOpen
  \bibfield  {author} {\bibinfo {author} {\bibfnamefont {A.~S.}\ \bibnamefont {Hernandez}}, \bibinfo {author} {\bibfnamefont {D.}~\bibnamefont {Moreno}}, \bibinfo {author} {\bibfnamefont {H.}~\bibnamefont {Bartosik}}, \bibinfo {author} {\bibfnamefont {N.}~\bibnamefont {Biancacci}}, \bibinfo {author} {\bibfnamefont {S.}~\bibnamefont {Hirlander}},\ and\ \bibinfo {author} {\bibfnamefont {A.}~\bibnamefont {Huschauer}},\ }\bibfield  {title} {{\selectlanguage {en}\bibinfo {title} {Space {Charge} studies on {LEIR}}},\ }\href {https://doi.org/10.1088/1742-6596/1067/6/062020} {\bibfield  {journal} {\bibinfo  {journal} {Journal of Physics: Conference Series}\ }\textbf {\bibinfo {volume} {1067}},\ \bibinfo {pages} {062020} (\bibinfo {year} {2018})},\ \bibinfo {note} {publisher: IOP Publishing}\BibitemShut {NoStop}%
\bibitem [{\citenamefont {Hirlaender}\ \emph {et~al.}(2018)\citenamefont {Hirlaender} \emph {et~al.}}]{hirlaender_lifetime_2018}%
  \BibitemOpen
  \bibfield  {author} {\bibinfo {author} {\bibfnamefont {S.}~\bibnamefont {Hirlaender}} \emph {et~al.},\ }\bibfield  {title} {{\selectlanguage {en}\bibinfo {title} {Lifetime and {Beam} {Losses} {Studies} of {Partially} {Strip} {Ions} in the {SPS} ({$^{129}$Xe}$^{39+}$)}},\ }\href {https://doi.org/10.18429/JACOW-IPAC2018-THPMF015} {\bibfield  {journal} {\bibinfo  {journal} {Proceedings of the 9th Int. Particle Accelerator Conf.}\ }\textbf {\bibinfo {volume} {IPAC2018}},\ \bibinfo {pages} {3 pages, 3.305 MB} (\bibinfo {year} {2018})},\ \bibinfo {note} {artwork Size: 3 pages, 3.305 MB ISBN: 9783954501847 Publisher: JACoW Publishing, Geneva, Switzerland}\BibitemShut {NoStop}%
\bibitem [{\citenamefont {Mahner}(2007)}]{mahner2007_LEIR_vacuum_system}%
  \BibitemOpen
  \bibfield  {author} {\bibinfo {author} {\bibfnamefont {E.}~\bibnamefont {Mahner}},\ }\bibfield  {title} {\bibinfo {title} {The vacuum system of the {Low Energy Ion Ring at CERN}: Requirements, design, and challenges},\ }\href@noop {} {\bibfield  {journal} {\bibinfo  {journal} {Vacuum}\ }\textbf {\bibinfo {volume} {81}},\ \bibinfo {pages} {727} (\bibinfo {year} {2007})}\BibitemShut {NoStop}%
\bibitem [{\citenamefont {Arduini}\ \emph {et~al.}(1996)\citenamefont {Arduini}, \citenamefont {Bailey}, \citenamefont {Bohl}, \citenamefont {Burkhardt}, \citenamefont {Cappi}, \citenamefont {Carter}, \citenamefont {Cornelis}, \citenamefont {Dach}, \citenamefont {de~Rijk}, \citenamefont {Faugier}, \citenamefont {Ferioli}, \citenamefont {Jakob}, \citenamefont {Jonker}, \citenamefont {Manglunki}, \citenamefont {Martini}, \citenamefont {Martini}, \citenamefont {Riunaud}, \citenamefont {Scheidenberger}, \citenamefont {Vandorpe}, \citenamefont {Vos},\ and\ \citenamefont {Zanolli}}]{arduini_stripper_foil_1996}%
  \BibitemOpen
  \bibfield  {author} {\bibinfo {author} {\bibfnamefont {G.}~\bibnamefont {Arduini}}, \bibinfo {author} {\bibfnamefont {R.}~\bibnamefont {Bailey}}, \bibinfo {author} {\bibfnamefont {T.}~\bibnamefont {Bohl}}, \bibinfo {author} {\bibfnamefont {H.}~\bibnamefont {Burkhardt}}, \bibinfo {author} {\bibfnamefont {R.}~\bibnamefont {Cappi}}, \bibinfo {author} {\bibfnamefont {C.}~\bibnamefont {Carter}}, \bibinfo {author} {\bibfnamefont {K.}~\bibnamefont {Cornelis}}, \bibinfo {author} {\bibfnamefont {M.}~\bibnamefont {Dach}}, \bibinfo {author} {\bibfnamefont {G.}~\bibnamefont {de~Rijk}}, \bibinfo {author} {\bibfnamefont {A.}~\bibnamefont {Faugier}}, \bibinfo {author} {\bibfnamefont {G.}~\bibnamefont {Ferioli}}, \bibinfo {author} {\bibfnamefont {H.}~\bibnamefont {Jakob}}, \bibinfo {author} {\bibfnamefont {M.}~\bibnamefont {Jonker}}, \bibinfo {author} {\bibfnamefont {D.}~\bibnamefont {Manglunki}}, \bibinfo {author} {\bibfnamefont {G.}~\bibnamefont {Martini}}, \bibinfo {author} {\bibfnamefont {M.}~\bibnamefont {Martini}},
  \bibinfo {author} {\bibfnamefont {J.~P.}\ \bibnamefont {Riunaud}}, \bibinfo {author} {\bibfnamefont {C.}~\bibnamefont {Scheidenberger}}, \bibinfo {author} {\bibfnamefont {B.}~\bibnamefont {Vandorpe}}, \bibinfo {author} {\bibfnamefont {L.}~\bibnamefont {Vos}},\ and\ \bibinfo {author} {\bibfnamefont {M.}~\bibnamefont {Zanolli}},\ }\bibfield  {title} {\bibinfo {title} {{Lead ion beam emittance and transmission studies in the {PS-SPS complex at CERN}}},\ }\href {https://cds.cern.ch/record/308372} {\  (\bibinfo {year} {1996})}\BibitemShut {NoStop}%
\bibitem [{\citenamefont {Kröger}\ \emph {et~al.}(2022)\citenamefont {Kröger}, \citenamefont {Weber}, \citenamefont {Hirlaender}, \citenamefont {Alemany-Fernandez}, \citenamefont {Krasny}, \citenamefont {Stöhlker}, \citenamefont {Tolstikhina},\ and\ \citenamefont {Shevelko}}]{kroeger_stripper_foil_2022}%
  \BibitemOpen
  \bibfield  {author} {\bibinfo {author} {\bibfnamefont {F.~M.}\ \bibnamefont {Kröger}}, \bibinfo {author} {\bibfnamefont {G.}~\bibnamefont {Weber}}, \bibinfo {author} {\bibfnamefont {S.}~\bibnamefont {Hirlaender}}, \bibinfo {author} {\bibfnamefont {R.}~\bibnamefont {Alemany-Fernandez}}, \bibinfo {author} {\bibfnamefont {M.~W.}\ \bibnamefont {Krasny}}, \bibinfo {author} {\bibfnamefont {T.}~\bibnamefont {Stöhlker}}, \bibinfo {author} {\bibfnamefont {I.~Y.}\ \bibnamefont {Tolstikhina}},\ and\ \bibinfo {author} {\bibfnamefont {V.~P.}\ \bibnamefont {Shevelko}},\ }\bibfield  {title} {{\selectlanguage {en}\bibinfo {title} {Charge-{State} {Distributions} of {Highly} {Charged} {Lead} {Ions} at {Relativistic} {Collision} {Energies}}},\ }\href {https://doi.org/10.1002/andp.202100245} {\bibfield  {journal} {\bibinfo  {journal} {Annalen der Physik}\ }\textbf {\bibinfo {volume} {534}},\ \bibinfo {pages} {2100245} (\bibinfo {year} {2022})}\BibitemShut {NoStop}%
\bibitem [{\citenamefont {Weber}\ \emph {et~al.}(2015)\citenamefont {Weber}, \citenamefont {Herdrich}, \citenamefont {DuBois}, \citenamefont {Hillenbrand}, \citenamefont {Beyer}, \citenamefont {Bozyk}, \citenamefont {Gassner}, \citenamefont {Grisenti}, \citenamefont {Hagmann}, \citenamefont {Litvinov} \emph {et~al.}}]{weber2015_total_EL_cross_section}%
  \BibitemOpen
  \bibfield  {author} {\bibinfo {author} {\bibfnamefont {G.}~\bibnamefont {Weber}}, \bibinfo {author} {\bibfnamefont {M.}~\bibnamefont {Herdrich}}, \bibinfo {author} {\bibfnamefont {R.}~\bibnamefont {DuBois}}, \bibinfo {author} {\bibfnamefont {P.-M.}\ \bibnamefont {Hillenbrand}}, \bibinfo {author} {\bibfnamefont {H.}~\bibnamefont {Beyer}}, \bibinfo {author} {\bibfnamefont {L.}~\bibnamefont {Bozyk}}, \bibinfo {author} {\bibfnamefont {T.}~\bibnamefont {Gassner}}, \bibinfo {author} {\bibfnamefont {R.}~\bibnamefont {Grisenti}}, \bibinfo {author} {\bibfnamefont {S.}~\bibnamefont {Hagmann}}, \bibinfo {author} {\bibfnamefont {Y.~A.}\ \bibnamefont {Litvinov}}, \emph {et~al.},\ }\bibfield  {title} {\bibinfo {title} {Total projectile electron loss cross sections of {U}$^{28+}$ ions in collisions with gaseous targets ranging from hydrogen to krypton},\ }\href@noop {} {\bibfield  {journal} {\bibinfo  {journal} {Physical Review Special Topics-Accelerators and Beams}\ }\textbf {\bibinfo {volume} {18}},\ \bibinfo {pages}
  {034403} (\bibinfo {year} {2015})}\BibitemShut {NoStop}%
\bibitem [{\citenamefont {Shevelko}\ \emph {et~al.}(2018)\citenamefont {Shevelko}, \citenamefont {Litvinov}, \citenamefont {St{\"o}hlker},\ and\ \citenamefont {Tolstikhina}}]{shevelko2018lifetimes_ricode_m}%
  \BibitemOpen
  \bibfield  {author} {\bibinfo {author} {\bibfnamefont {V.}~\bibnamefont {Shevelko}}, \bibinfo {author} {\bibfnamefont {Y.~A.}\ \bibnamefont {Litvinov}}, \bibinfo {author} {\bibfnamefont {T.}~\bibnamefont {St{\"o}hlker}},\ and\ \bibinfo {author} {\bibfnamefont {I.~Y.}\ \bibnamefont {Tolstikhina}},\ }\bibfield  {title} {\bibinfo {title} {Lifetimes of relativistic heavy-ion beams in the high energy storage ring of {FAIR}},\ }\href@noop {} {\bibfield  {journal} {\bibinfo  {journal} {Nuclear Instruments and Methods in Physics Research Section B: Beam Interactions with Materials and Atoms}\ }\textbf {\bibinfo {volume} {421}},\ \bibinfo {pages} {45} (\bibinfo {year} {2018})}\BibitemShut {NoStop}%
\bibitem [{\citenamefont {Tolstikhina}\ \emph {et~al.}(2018)\citenamefont {Tolstikhina}, \citenamefont {Imai}, \citenamefont {Winckler},\ and\ \citenamefont {Shevelko}}]{basic_atomic_interactions_book_tolstikhina2018}%
  \BibitemOpen
  \bibfield  {author} {\bibinfo {author} {\bibfnamefont {I.}~\bibnamefont {Tolstikhina}}, \bibinfo {author} {\bibfnamefont {M.}~\bibnamefont {Imai}}, \bibinfo {author} {\bibfnamefont {N.}~\bibnamefont {Winckler}},\ and\ \bibinfo {author} {\bibfnamefont {V.}~\bibnamefont {Shevelko}},\ }\href@noop {} {\emph {\bibinfo {title} {Basic Atomic Interactions of Accelerated Heavy Ions in Matter}}}\ (\bibinfo  {publisher} {Springer},\ \bibinfo {year} {2018})\BibitemShut {NoStop}%
\bibitem [{\citenamefont {Shevelko}\ \emph {et~al.}(2011)\citenamefont {Shevelko}, \citenamefont {Beigman}, \citenamefont {Litsarev}, \citenamefont {Tawara}, \citenamefont {Tolstikhina},\ and\ \citenamefont {Weber}}]{shevelko2011_summary}%
  \BibitemOpen
  \bibfield  {author} {\bibinfo {author} {\bibfnamefont {V.}~\bibnamefont {Shevelko}}, \bibinfo {author} {\bibfnamefont {I.}~\bibnamefont {Beigman}}, \bibinfo {author} {\bibfnamefont {M.}~\bibnamefont {Litsarev}}, \bibinfo {author} {\bibfnamefont {H.}~\bibnamefont {Tawara}}, \bibinfo {author} {\bibfnamefont {I.~Y.}\ \bibnamefont {Tolstikhina}},\ and\ \bibinfo {author} {\bibfnamefont {G.}~\bibnamefont {Weber}},\ }\bibfield  {title} {\bibinfo {title} {Charge-changing processes in collisions of heavy many-electron ions with neutral atoms},\ }\href@noop {} {\bibfield  {journal} {\bibinfo  {journal} {Nuclear Instruments and Methods in Physics Research Section B: Beam Interactions with Materials and Atoms}\ }\textbf {\bibinfo {volume} {269}},\ \bibinfo {pages} {1455} (\bibinfo {year} {2011})}\BibitemShut {NoStop}%
\bibitem [{\citenamefont {Citron}\ \emph {et~al.}(2018)\citenamefont {Citron} \emph {et~al.}}]{citron_arxiv_2018}%
  \BibitemOpen
  \bibfield  {author} {\bibinfo {author} {\bibfnamefont {Z.}~\bibnamefont {Citron}} \emph {et~al.},\ }\href {https://doi.org/10.23731/CYRM-2019-007.1159} {{\selectlanguage {en}\bibinfo {title} {{arXiv} : {Report} from {Working} {Group} 5 : {Future} physics opportunities for high-density {QCD} at the {LHC} with heavy-ion and proton beams}}} (\bibinfo {year} {2018}),\ \bibinfo {note} {number: CERN-LPCC-2018-07}\BibitemShut {NoStop}%
\bibitem [{\citenamefont {{ALICE Collaboration}}(2022)}]{alice_collaboration_letter_2022}%
  \BibitemOpen
  \bibfield  {author} {\bibinfo {author} {\bibnamefont {{ALICE Collaboration}}},\ }\href {http://arxiv.org/abs/2211.02491} {\bibinfo {title} {Letter of intent for {ALICE} 3: {A} next-generation heavy-ion experiment at the {LHC}}} (\bibinfo {year} {2022}),\ \bibinfo {note} {arXiv:2211.02491 [hep-ex, physics:nucl-ex, physics:physics]}\BibitemShut {NoStop}%
\bibitem [{\citenamefont {Mackowiak-Pawlowska}(2023)}]{mackowiak-pawlowska_addendum_2023}%
  \BibitemOpen
  \bibfield  {author} {\bibinfo {author} {\bibfnamefont {M.}~\bibnamefont {Mackowiak-Pawlowska}},\ }\href {https://cds.cern.ch/record/2867952} {\bibinfo {title} {Addendum to the {NA61}/{SHINE} {Proposal}: {Request} for light ions beams in {Run} 4}} (\bibinfo {year} {2023}),\ \bibinfo {note} {report No. CERN-SPSC-2023-022}\BibitemShut {NoStop}%
\bibitem [{\citenamefont {Bruce}\ \emph {et~al.}(2021)\citenamefont {Bruce}, \citenamefont {Jowett}, \citenamefont {Alemany-Fernández}, \citenamefont {Jebramcik}, \citenamefont {Schaumann},\ and\ \citenamefont {Bartosik}}]{bruce_O_pliot_studies_2021}%
  \BibitemOpen
  \bibfield  {author} {\bibinfo {author} {\bibfnamefont {R.}~\bibnamefont {Bruce}}, \bibinfo {author} {\bibfnamefont {J.}~\bibnamefont {Jowett}}, \bibinfo {author} {\bibfnamefont {R.}~\bibnamefont {Alemany-Fernández}}, \bibinfo {author} {\bibfnamefont {M.}~\bibnamefont {Jebramcik}}, \bibinfo {author} {\bibfnamefont {M.}~\bibnamefont {Schaumann}},\ and\ \bibinfo {author} {\bibfnamefont {H.}~\bibnamefont {Bartosik}},\ }\href {https://doi.org/10.18429/JACoW-IPAC2021-MOPAB005} {{\selectlanguage {sv}\bibinfo {title} {Studies for an {LHC} {Pilot} {Run} with {Oxygen} {Beams}}}} (\bibinfo {year} {2021}),\ \bibinfo {note} {10.18429/JACoW-IPAC2021-MOPAB005. Available at \url{https://cds.cern.ch/record/2783801}}\BibitemShut {NoStop}%
\bibitem [{\citenamefont {Browzet}\ and\ \citenamefont {Middlekoop}(1987)}]{browzet_performance_1987_first_oxygen}%
  \BibitemOpen
  \bibfield  {author} {\bibinfo {author} {\bibfnamefont {E.}~\bibnamefont {Browzet}}\ and\ \bibinfo {author} {\bibfnamefont {W.}~\bibnamefont {Middlekoop}},\ }\href@noop {} {\emph {\bibinfo {title} {Performance of the {PS} and {SPS} accelerator complex with oxygen ions}}}\ (\bibinfo  {publisher} {IEEE Service Center},\ \bibinfo {address} {United States},\ \bibinfo {year} {1987})\ \bibinfo {note} {iNIS Reference Number: 19084034}\BibitemShut {NoStop}%
\bibitem [{\citenamefont {Franzke}(1992)}]{Franzke_1992_interaction}%
  \BibitemOpen
  \bibfield  {author} {\bibinfo {author} {\bibfnamefont {B.}~\bibnamefont {Franzke}},\ }\bibfield  {title} {\bibinfo {title} {{Interaction of stored ion beams with the residual gas}},\ }in\ \href {https://doi.org/10.5170/CERN-1992-001.100} {\emph {\bibinfo {booktitle} {CERN Yellow Reports: School Proceedings}}}\ (\bibinfo {year} {1992})\BibitemShut {NoStop}%
\bibitem [{\citenamefont {Zimmermann}(1993)}]{zimmermann1993emittance_HERA}%
  \BibitemOpen
  \bibfield  {author} {\bibinfo {author} {\bibfnamefont {F.}~\bibnamefont {Zimmermann}},\ }\emph {\bibinfo {title} {Emittance growth and proton beam lifetime in {HERA}}},\ \href@noop {} {Ph.D. thesis},\ \bibinfo  {school} {Universit{\"a}t Hamburg} (\bibinfo {year} {1993})\BibitemShut {NoStop}%
\bibitem [{\citenamefont {Luo}\ \emph {et~al.}(2016)\citenamefont {Luo}, \citenamefont {Fischer},\ and\ \citenamefont {White}}]{rhic_proton_lifetimes_luo2016}%
  \BibitemOpen
  \bibfield  {author} {\bibinfo {author} {\bibfnamefont {Y.}~\bibnamefont {Luo}}, \bibinfo {author} {\bibfnamefont {W.}~\bibnamefont {Fischer}},\ and\ \bibinfo {author} {\bibfnamefont {S.}~\bibnamefont {White}},\ }\bibfield  {title} {\bibinfo {title} {Analysis and modeling of proton beam loss and emittance growth in the {Relativistic Heavy Ion Collider}},\ }\href@noop {} {\bibfield  {journal} {\bibinfo  {journal} {Physical Review Accelerators and Beams}\ }\textbf {\bibinfo {volume} {19}},\ \bibinfo {pages} {021001} (\bibinfo {year} {2016})}\BibitemShut {NoStop}%
\bibitem [{\citenamefont {Olson}\ \emph {et~al.}(1988)\citenamefont {Olson}, \citenamefont {Wang},\ and\ \citenamefont {Ullrich}}]{olson1988electronic_EC_classical_trajector_MC}%
  \BibitemOpen
  \bibfield  {author} {\bibinfo {author} {\bibfnamefont {R.}~\bibnamefont {Olson}}, \bibinfo {author} {\bibfnamefont {J.}~\bibnamefont {Wang}},\ and\ \bibinfo {author} {\bibfnamefont {J.}~\bibnamefont {Ullrich}},\ }\href@noop {} {\bibinfo {title} {Electronic and atomic collisions ed hb gilbody, wr newell, fh read and ach smith}} (\bibinfo {year} {1988})\BibitemShut {NoStop}%
\bibitem [{\citenamefont {Eichler}(1981)}]{eichler1981_eikonal}%
  \BibitemOpen
  \bibfield  {author} {\bibinfo {author} {\bibfnamefont {J.~K.}\ \bibnamefont {Eichler}},\ }\bibfield  {title} {\bibinfo {title} {Eikonal theory of charge exchange between arbitrary hydrogenic states of target and projectile},\ }\href@noop {} {\bibfield  {journal} {\bibinfo  {journal} {Physical Review A}\ }\textbf {\bibinfo {volume} {23}},\ \bibinfo {pages} {498} (\bibinfo {year} {1981})}\BibitemShut {NoStop}%
\bibitem [{\citenamefont {Shevelko}\ \emph {et~al.}(2003)\citenamefont {Shevelko}, \citenamefont {Rosmej}, \citenamefont {Tawara},\ and\ \citenamefont {Tolstikhina}}]{shevelko2003target_capture_code}%
  \BibitemOpen
  \bibfield  {author} {\bibinfo {author} {\bibfnamefont {V.}~\bibnamefont {Shevelko}}, \bibinfo {author} {\bibfnamefont {O.}~\bibnamefont {Rosmej}}, \bibinfo {author} {\bibfnamefont {H.}~\bibnamefont {Tawara}},\ and\ \bibinfo {author} {\bibfnamefont {I.~Y.}\ \bibnamefont {Tolstikhina}},\ }\bibfield  {title} {\bibinfo {title} {The target-density effect in electron-capture processes},\ }\href@noop {} {\bibfield  {journal} {\bibinfo  {journal} {Journal of Physics B: Atomic, Molecular and Optical Physics}\ }\textbf {\bibinfo {volume} {37}},\ \bibinfo {pages} {201} (\bibinfo {year} {2003})}\BibitemShut {NoStop}%
\bibitem [{\citenamefont {Schlachter}\ \emph {et~al.}(1983)\citenamefont {Schlachter}, \citenamefont {Stearns}, \citenamefont {Graham}, \citenamefont {Berkner}, \citenamefont {Pyle},\ and\ \citenamefont {Tanis}}]{schlachter_electron_1983}%
  \BibitemOpen
  \bibfield  {author} {\bibinfo {author} {\bibfnamefont {A.~S.}\ \bibnamefont {Schlachter}}, \bibinfo {author} {\bibfnamefont {J.~W.}\ \bibnamefont {Stearns}}, \bibinfo {author} {\bibfnamefont {W.~G.}\ \bibnamefont {Graham}}, \bibinfo {author} {\bibfnamefont {K.~H.}\ \bibnamefont {Berkner}}, \bibinfo {author} {\bibfnamefont {R.~V.}\ \bibnamefont {Pyle}},\ and\ \bibinfo {author} {\bibfnamefont {J.~A.}\ \bibnamefont {Tanis}},\ }\bibfield  {title} {{\selectlanguage {en}\bibinfo {title} {Electron capture for fast highly charged ions in gas targets: {An} empirical scaling rule}},\ }\href {https://doi.org/10.1103/PhysRevA.27.3372} {\bibfield  {journal} {\bibinfo  {journal} {Physical Review A}\ }\textbf {\bibinfo {volume} {27}},\ \bibinfo {pages} {3372} (\bibinfo {year} {1983})}\BibitemShut {NoStop}%
\bibitem [{\citenamefont {Shevelko}\ \emph {et~al.}(2010)\citenamefont {Shevelko}, \citenamefont {Stöhlker}, \citenamefont {Tawara}, \citenamefont {Tolstikhina},\ and\ \citenamefont {Weber}}]{shevelko_electron_2010}%
  \BibitemOpen
  \bibfield  {author} {\bibinfo {author} {\bibfnamefont {V.~P.}\ \bibnamefont {Shevelko}}, \bibinfo {author} {\bibfnamefont {T.}~\bibnamefont {Stöhlker}}, \bibinfo {author} {\bibfnamefont {H.}~\bibnamefont {Tawara}}, \bibinfo {author} {\bibfnamefont {I.~Y.}\ \bibnamefont {Tolstikhina}},\ and\ \bibinfo {author} {\bibfnamefont {G.}~\bibnamefont {Weber}},\ }\bibfield  {title} {{\selectlanguage {en}\bibinfo {title} {Electron capture in intermediate-to-fast heavy ion collisions with neutral atoms}},\ }\href {https://doi.org/10.1016/j.nimb.2010.06.019} {\bibfield  {journal} {\bibinfo  {journal} {Nuclear Instruments and Methods in Physics Research Section B: Beam Interactions with Materials and Atoms}\ }\textbf {\bibinfo {volume} {268}},\ \bibinfo {pages} {2611} (\bibinfo {year} {2010})}\BibitemShut {NoStop}%
\bibitem [{\citenamefont {Baur}\ \emph {et~al.}(2009)\citenamefont {Baur}, \citenamefont {Beigman}, \citenamefont {Shevelko}, \citenamefont {Tolstikhina},\ and\ \citenamefont {St{\"o}hlker}}]{baur2009ionization_EL_data_and_Born_approx}%
  \BibitemOpen
  \bibfield  {author} {\bibinfo {author} {\bibfnamefont {G.}~\bibnamefont {Baur}}, \bibinfo {author} {\bibfnamefont {I.}~\bibnamefont {Beigman}}, \bibinfo {author} {\bibfnamefont {V.}~\bibnamefont {Shevelko}}, \bibinfo {author} {\bibfnamefont {I.~Y.}\ \bibnamefont {Tolstikhina}},\ and\ \bibinfo {author} {\bibfnamefont {T.}~\bibnamefont {St{\"o}hlker}},\ }\bibfield  {title} {\bibinfo {title} {Ionization of highly charged relativistic ions by neutral atoms and ions},\ }\href@noop {} {\bibfield  {journal} {\bibinfo  {journal} {Physical Review A}\ }\textbf {\bibinfo {volume} {80}},\ \bibinfo {pages} {012713} (\bibinfo {year} {2009})}\BibitemShut {NoStop}%
\bibitem [{\citenamefont {Shevelko}\ \emph {et~al.}(2001)\citenamefont {Shevelko}, \citenamefont {Tolstikhina},\ and\ \citenamefont {St{\"o}hlker}}]{loss}%
  \BibitemOpen
  \bibfield  {author} {\bibinfo {author} {\bibfnamefont {V.}~\bibnamefont {Shevelko}}, \bibinfo {author} {\bibfnamefont {I.~Y.}\ \bibnamefont {Tolstikhina}},\ and\ \bibinfo {author} {\bibfnamefont {T.}~\bibnamefont {St{\"o}hlker}},\ }\bibfield  {title} {\bibinfo {title} {Stripping of fast heavy low-charged ions in gaseous targets},\ }\href@noop {} {\bibfield  {journal} {\bibinfo  {journal} {Nuclear Instruments and Methods in Physics Research Section B: Beam Interactions with Materials and Atoms}\ }\textbf {\bibinfo {volume} {184}},\ \bibinfo {pages} {295} (\bibinfo {year} {2001})}\BibitemShut {NoStop}%
\bibitem [{\citenamefont {Beigman}\ \emph {et~al.}(2008)\citenamefont {Beigman}, \citenamefont {Tolstikhina},\ and\ \citenamefont {Shevelko}}]{loss-r}%
  \BibitemOpen
  \bibfield  {author} {\bibinfo {author} {\bibfnamefont {I.}~\bibnamefont {Beigman}}, \bibinfo {author} {\bibfnamefont {I.~Y.}\ \bibnamefont {Tolstikhina}},\ and\ \bibinfo {author} {\bibfnamefont {V.}~\bibnamefont {Shevelko}},\ }\bibfield  {title} {\bibinfo {title} {Ionization of heavy ions in relativistic collisions with neutral atoms},\ }\href@noop {} {\bibfield  {journal} {\bibinfo  {journal} {Technical Physics}\ }\textbf {\bibinfo {volume} {53}},\ \bibinfo {pages} {547} (\bibinfo {year} {2008})}\BibitemShut {NoStop}%
\bibitem [{\citenamefont {Tolstikhina}\ \emph {et~al.}(2014{\natexlab{a}})\citenamefont {Tolstikhina}, \citenamefont {Tupitsyn}, \citenamefont {Andreev},\ and\ \citenamefont {Shevelko}}]{tolstikhina2014influence_el_RICODE}%
  \BibitemOpen
  \bibfield  {author} {\bibinfo {author} {\bibfnamefont {I.~Y.}\ \bibnamefont {Tolstikhina}}, \bibinfo {author} {\bibfnamefont {I.}~\bibnamefont {Tupitsyn}}, \bibinfo {author} {\bibfnamefont {S.}~\bibnamefont {Andreev}},\ and\ \bibinfo {author} {\bibfnamefont {V.}~\bibnamefont {Shevelko}},\ }\bibfield  {title} {\bibinfo {title} {Influence of relativistic effects on electron-loss cross sections of heavy and superheavy ions colliding with neutral atoms},\ }\href@noop {} {\bibfield  {journal} {\bibinfo  {journal} {Journal of Experimental and Theoretical Physics}\ }\textbf {\bibinfo {volume} {119}},\ \bibinfo {pages} {1} (\bibinfo {year} {2014}{\natexlab{a}})}\BibitemShut {NoStop}%
\bibitem [{\citenamefont {Tolstikhina}\ \emph {et~al.}(2014{\natexlab{b}})\citenamefont {Tolstikhina}, \citenamefont {Tupitsyn}, \citenamefont {Andreev},\ and\ \citenamefont {Shevelko}}]{tolstikhina2014influence_ricode_m_first_suggested}%
  \BibitemOpen
  \bibfield  {author} {\bibinfo {author} {\bibfnamefont {I.~Y.}\ \bibnamefont {Tolstikhina}}, \bibinfo {author} {\bibfnamefont {I.}~\bibnamefont {Tupitsyn}}, \bibinfo {author} {\bibfnamefont {S.}~\bibnamefont {Andreev}},\ and\ \bibinfo {author} {\bibfnamefont {V.}~\bibnamefont {Shevelko}},\ }\bibfield  {title} {\bibinfo {title} {Influence of relativistic effects on electron-loss cross sections of heavy and superheavy ions colliding with neutral atoms},\ }\href@noop {} {\bibfield  {journal} {\bibinfo  {journal} {Journal of Experimental and Theoretical Physics}\ }\textbf {\bibinfo {volume} {119}},\ \bibinfo {pages} {1} (\bibinfo {year} {2014}{\natexlab{b}})}\BibitemShut {NoStop}%
\bibitem [{\citenamefont {Ralchenko}(2005)}]{nist_database}%
  \BibitemOpen
  \bibfield  {author} {\bibinfo {author} {\bibfnamefont {Y.}~\bibnamefont {Ralchenko}},\ }\bibfield  {title} {\bibinfo {title} {{NIST} atomic spectra database},\ }\href@noop {} {\bibfield  {journal} {\bibinfo  {journal} {Memorie della Societ{\`a} Astronomica Italiana Supplement, v. 8, p. 96 (2005)}\ }\textbf {\bibinfo {volume} {8}},\ \bibinfo {pages} {96} (\bibinfo {year} {2005})},\ \bibinfo {note} {{Ionization Energies Form} available at \url{https://physics.nist.gov/PhysRefData/ASD/ionEnergy.html}}\BibitemShut {NoStop}%
\bibitem [{\citenamefont {DuBois}\ \emph {et~al.}(2011)\citenamefont {DuBois}, \citenamefont {Santos}, \citenamefont {Sigaud},\ and\ \citenamefont {Montenegro}}]{dubois_electron_2011}%
  \BibitemOpen
  \bibfield  {author} {\bibinfo {author} {\bibfnamefont {R.~D.}\ \bibnamefont {DuBois}}, \bibinfo {author} {\bibfnamefont {A.~C.~F.}\ \bibnamefont {Santos}}, \bibinfo {author} {\bibfnamefont {G.~M.}\ \bibnamefont {Sigaud}},\ and\ \bibinfo {author} {\bibfnamefont {E.~C.}\ \bibnamefont {Montenegro}},\ }\bibfield  {title} {\bibinfo {title} {Electron loss from fast heavy ions: {Target}-scaling dependence},\ }\href {https://doi.org/10.1103/PhysRevA.84.022702} {\bibfield  {journal} {\bibinfo  {journal} {Physical Review A}\ }\textbf {\bibinfo {volume} {84}},\ \bibinfo {pages} {022702} (\bibinfo {year} {2011})},\ \bibinfo {note} {publisher: American Physical Society}\BibitemShut {NoStop}%
\bibitem [{\citenamefont {Weber}(2017)}]{weber_2016_semi_empirical_formula}%
  \BibitemOpen
  \bibfield  {author} {\bibinfo {author} {\bibfnamefont {G.}~\bibnamefont {Weber}},\ }\href@noop {} {\bibinfo {title} {A new semi-empirical formular for total electron loss by energetic ions}} (\bibinfo {year} {2017}),\ \bibinfo {note} {2016 Annual Report Helmholtz-Institut Jena. DOI: 10.15120/GSI-2017-00708. Available at \url{http://repository.gsi.de/record/201624}.}\BibitemShut {Stop}%
\bibitem [{\citenamefont {H{\"u}lsk{\"o}tter}\ \emph {et~al.}(1991)\citenamefont {H{\"u}lsk{\"o}tter}, \citenamefont {Feinberg}, \citenamefont {Meyerhof}, \citenamefont {Belkacem}, \citenamefont {Alonso}, \citenamefont {Blumenfeld}, \citenamefont {Dillard}, \citenamefont {Gould}, \citenamefont {Guardala}, \citenamefont {Krebs} \emph {et~al.}}]{hulskotter1991electron_Au52_EL_benchmark}%
  \BibitemOpen
  \bibfield  {author} {\bibinfo {author} {\bibfnamefont {H.-P.}\ \bibnamefont {H{\"u}lsk{\"o}tter}}, \bibinfo {author} {\bibfnamefont {B.}~\bibnamefont {Feinberg}}, \bibinfo {author} {\bibfnamefont {W.}~\bibnamefont {Meyerhof}}, \bibinfo {author} {\bibfnamefont {A.}~\bibnamefont {Belkacem}}, \bibinfo {author} {\bibfnamefont {J.}~\bibnamefont {Alonso}}, \bibinfo {author} {\bibfnamefont {L.}~\bibnamefont {Blumenfeld}}, \bibinfo {author} {\bibfnamefont {E.}~\bibnamefont {Dillard}}, \bibinfo {author} {\bibfnamefont {H.}~\bibnamefont {Gould}}, \bibinfo {author} {\bibfnamefont {N.}~\bibnamefont {Guardala}}, \bibinfo {author} {\bibfnamefont {G.}~\bibnamefont {Krebs}}, \emph {et~al.},\ }\bibfield  {title} {\bibinfo {title} {Electron-electron interaction in projectile electron loss},\ }\href@noop {} {\bibfield  {journal} {\bibinfo  {journal} {Physical Review A}\ }\textbf {\bibinfo {volume} {44}},\ \bibinfo {pages} {1712} (\bibinfo {year} {1991})}\BibitemShut {NoStop}%
\bibitem [{\citenamefont {Alton}\ \emph {et~al.}(1981)\citenamefont {Alton}, \citenamefont {Bridwell}, \citenamefont {Lucas}, \citenamefont {Moak}, \citenamefont {Miller}, \citenamefont {Jones}, \citenamefont {Kessel}, \citenamefont {Antar},\ and\ \citenamefont {Brown}}]{alton1981single_Fe4_EL_benchmark}%
  \BibitemOpen
  \bibfield  {author} {\bibinfo {author} {\bibfnamefont {G.}~\bibnamefont {Alton}}, \bibinfo {author} {\bibfnamefont {L.}~\bibnamefont {Bridwell}}, \bibinfo {author} {\bibfnamefont {M.}~\bibnamefont {Lucas}}, \bibinfo {author} {\bibfnamefont {C.}~\bibnamefont {Moak}}, \bibinfo {author} {\bibfnamefont {P.}~\bibnamefont {Miller}}, \bibinfo {author} {\bibfnamefont {C.}~\bibnamefont {Jones}}, \bibinfo {author} {\bibfnamefont {Q.}~\bibnamefont {Kessel}}, \bibinfo {author} {\bibfnamefont {A.}~\bibnamefont {Antar}},\ and\ \bibinfo {author} {\bibfnamefont {M.}~\bibnamefont {Brown}},\ }\bibfield  {title} {\bibinfo {title} {Single-and multiple-electron-loss-cross-section measurements from {20-MeV Fe 4+} on thin gaseous targets},\ }\href@noop {} {\bibfield  {journal} {\bibinfo  {journal} {Physical Review A}\ }\textbf {\bibinfo {volume} {23}},\ \bibinfo {pages} {1073} (\bibinfo {year} {1981})}\BibitemShut {NoStop}%
\bibitem [{\citenamefont {Peng}(2004)}]{peng2004dependence_Xe18_EL_benchmark}%
  \BibitemOpen
  \bibfield  {author} {\bibinfo {author} {\bibfnamefont {Y.}~\bibnamefont {Peng}},\ }\emph {\bibinfo {title} {Dependence of cross sections for multi-electron loss by 6 {MeV/amu Xe18+} ions on target atomic number}},\ \href@noop {} {Ph.D. thesis},\ \bibinfo  {school} {Texas A\&M University} (\bibinfo {year} {2004})\BibitemShut {NoStop}%
\bibitem [{\citenamefont {Meyerhof}\ \emph {et~al.}(1987{\natexlab{a}})\citenamefont {Meyerhof}, \citenamefont {Anholt},\ and\ \citenamefont {Xu}}]{meyerhof1987atomic_Xe45_EL_benchmark}%
  \BibitemOpen
  \bibfield  {author} {\bibinfo {author} {\bibfnamefont {W.}~\bibnamefont {Meyerhof}}, \bibinfo {author} {\bibfnamefont {R.}~\bibnamefont {Anholt}},\ and\ \bibinfo {author} {\bibfnamefont {X.-Y.}\ \bibnamefont {Xu}},\ }\bibfield  {title} {\bibinfo {title} {Atomic collisions with relativistic heavy ions. vii. l-and m-shell electron stripping of ions in light targets},\ }\href@noop {} {\bibfield  {journal} {\bibinfo  {journal} {Physical Review A}\ }\textbf {\bibinfo {volume} {35}},\ \bibinfo {pages} {1055} (\bibinfo {year} {1987}{\natexlab{a}})}\BibitemShut {NoStop}%
\bibitem [{\citenamefont {Meyerhof}\ \emph {et~al.}(1987{\natexlab{b}})\citenamefont {Meyerhof}, \citenamefont {Anholt}, \citenamefont {Xu}, \citenamefont {Gould}, \citenamefont {Feinberg}, \citenamefont {McDonald}, \citenamefont {Wegner},\ and\ \citenamefont {Thieberger}}]{meyerhof1987multiple_U83_EL_benchmark}%
  \BibitemOpen
  \bibfield  {author} {\bibinfo {author} {\bibfnamefont {W.}~\bibnamefont {Meyerhof}}, \bibinfo {author} {\bibfnamefont {R.}~\bibnamefont {Anholt}}, \bibinfo {author} {\bibfnamefont {X.-Y.}\ \bibnamefont {Xu}}, \bibinfo {author} {\bibfnamefont {H.}~\bibnamefont {Gould}}, \bibinfo {author} {\bibfnamefont {B.}~\bibnamefont {Feinberg}}, \bibinfo {author} {\bibfnamefont {R.}~\bibnamefont {McDonald}}, \bibinfo {author} {\bibfnamefont {H.}~\bibnamefont {Wegner}},\ and\ \bibinfo {author} {\bibfnamefont {P.}~\bibnamefont {Thieberger}},\ }\bibfield  {title} {\bibinfo {title} {Multiple ionization in relativistic heavy-ion--atom collisions},\ }\href@noop {} {\bibfield  {journal} {\bibinfo  {journal} {Physical Review A}\ }\textbf {\bibinfo {volume} {35}},\ \bibinfo {pages} {1967} (\bibinfo {year} {1987}{\natexlab{b}})}\BibitemShut {NoStop}%
\bibitem [{bea()}]{beam_gas_collisions_github_repo}%
  \BibitemOpen
  \href@noop {} {\bibinfo {title} {Beam-gas collisions github repository}},\ \bibinfo {note} {available at: \url{https://github.com/ewaagaard/Beam-gas-collisions}}\BibitemShut {NoStop}%
\bibitem [{\citenamefont {Franzke}(1981)}]{franzke1981_semi_empirical_vacuum_formula}%
  \BibitemOpen
  \bibfield  {author} {\bibinfo {author} {\bibfnamefont {B.}~\bibnamefont {Franzke}},\ }\bibfield  {title} {\bibinfo {title} {Vacuum requirements for heavy ion synchrotrons},\ }\href@noop {} {\bibfield  {journal} {\bibinfo  {journal} {IEEE Transactions on Nuclear Science}\ }\textbf {\bibinfo {volume} {28}},\ \bibinfo {pages} {2116} (\bibinfo {year} {1981})}\BibitemShut {NoStop}%
\bibitem [{\citenamefont {Olson}\ \emph {et~al.}(2004)\citenamefont {Olson}, \citenamefont {Watson}, \citenamefont {Horvat}, \citenamefont {Perumal}, \citenamefont {Peng},\ and\ \citenamefont {St{\"o}hlker}}]{olson2004projectile_EL}%
  \BibitemOpen
  \bibfield  {author} {\bibinfo {author} {\bibfnamefont {R.}~\bibnamefont {Olson}}, \bibinfo {author} {\bibfnamefont {R.}~\bibnamefont {Watson}}, \bibinfo {author} {\bibfnamefont {V.}~\bibnamefont {Horvat}}, \bibinfo {author} {\bibfnamefont {A.}~\bibnamefont {Perumal}}, \bibinfo {author} {\bibfnamefont {Y.}~\bibnamefont {Peng}},\ and\ \bibinfo {author} {\bibfnamefont {T.}~\bibnamefont {St{\"o}hlker}},\ }\bibfield  {title} {\bibinfo {title} {Projectile electron loss and capture in {MeV/u collisions of U$^{28+}$ with H$_2$, N$_2$ and Ar}},\ }\href@noop {} {\bibfield  {journal} {\bibinfo  {journal} {Journal of Physics B: Atomic, Molecular and Optical Physics}\ }\textbf {\bibinfo {volume} {37}},\ \bibinfo {pages} {4539} (\bibinfo {year} {2004})}\BibitemShut {NoStop}%
\bibitem [{\citenamefont {Perumal}\ \emph {et~al.}(2005)\citenamefont {Perumal}, \citenamefont {Horvat}, \citenamefont {Watson}, \citenamefont {Peng},\ and\ \citenamefont {Fruchey}}]{perumal2005_multiple_EC}%
  \BibitemOpen
  \bibfield  {author} {\bibinfo {author} {\bibfnamefont {A.}~\bibnamefont {Perumal}}, \bibinfo {author} {\bibfnamefont {V.}~\bibnamefont {Horvat}}, \bibinfo {author} {\bibfnamefont {R.}~\bibnamefont {Watson}}, \bibinfo {author} {\bibfnamefont {Y.}~\bibnamefont {Peng}},\ and\ \bibinfo {author} {\bibfnamefont {K.}~\bibnamefont {Fruchey}},\ }\bibfield  {title} {\bibinfo {title} {Cross sections for charge change in argon and equilibrium charge states of 3.5 {MeV/amu} uranium ions passing through argon and carbon targets},\ }\href@noop {} {\bibfield  {journal} {\bibinfo  {journal} {Nuclear Instruments and Methods in Physics Research Section B: Beam Interactions with Materials and Atoms}\ }\textbf {\bibinfo {volume} {227}},\ \bibinfo {pages} {251} (\bibinfo {year} {2005})}\BibitemShut {NoStop}%
\bibitem [{\citenamefont {Erb}(1978)}]{erb1978_gsi_EC_data}%
  \BibitemOpen
  \bibfield  {author} {\bibinfo {author} {\bibfnamefont {W.}~\bibnamefont {Erb}},\ }\href {https://cds.cern.ch/record/119615} {\emph {\bibinfo {title} {{Umladung schwerer Ionen nach Durchgang durch Gase und Festkörper im Energiebereich 0,2 bis 1,4 MeV/u}}}}\ (\bibinfo  {publisher} {GSI},\ \bibinfo {address} {Darmstadt},\ \bibinfo {year} {1978})\BibitemShut {NoStop}%
\bibitem [{\citenamefont {Mandl}(1991)}]{mandl1991statistical_ideal_gas_law}%
  \BibitemOpen
  \bibfield  {author} {\bibinfo {author} {\bibfnamefont {F.}~\bibnamefont {Mandl}},\ }\href@noop {} {\emph {\bibinfo {title} {Statistical physics}}},\ Vol.~\bibinfo {volume} {14}\ (\bibinfo  {publisher} {John Wiley \& Sons},\ \bibinfo {year} {1991})\BibitemShut {NoStop}%
\bibitem [{\citenamefont {Dalton}(1802)}]{dalton1802essay}%
  \BibitemOpen
  \bibfield  {author} {\bibinfo {author} {\bibfnamefont {J.}~\bibnamefont {Dalton}},\ }\bibfield  {title} {\bibinfo {title} {{Essay IV. On the expansion of elastic fluids by heat}},\ }\href@noop {} {\bibfield  {journal} {\bibinfo  {journal} {Memoirs of the Literary and Philosophical Society of Manchester}\ }\textbf {\bibinfo {volume} {5}},\ \bibinfo {pages} {595} (\bibinfo {year} {1802})}\BibitemShut {NoStop}%
\bibitem [{\citenamefont {Watson}\ \emph {et~al.}(2003)\citenamefont {Watson}, \citenamefont {Peng}, \citenamefont {Horvat}, \citenamefont {Kim},\ and\ \citenamefont {Olson}}]{additivity_rule}%
  \BibitemOpen
  \bibfield  {author} {\bibinfo {author} {\bibfnamefont {R.~L.}\ \bibnamefont {Watson}}, \bibinfo {author} {\bibfnamefont {Y.}~\bibnamefont {Peng}}, \bibinfo {author} {\bibfnamefont {V.}~\bibnamefont {Horvat}}, \bibinfo {author} {\bibfnamefont {G.~J.}\ \bibnamefont {Kim}},\ and\ \bibinfo {author} {\bibfnamefont {R.~E.}\ \bibnamefont {Olson}},\ }\bibfield  {title} {\bibinfo {title} {Target {Z} dependence and additivity of cross sections for electron loss by 6-{MeV/amu Xe}$^{18+}$ projectiles},\ }\href {https://doi.org/10.1103/PhysRevA.67.022706} {\bibfield  {journal} {\bibinfo  {journal} {Phys. Rev. A}\ }\textbf {\bibinfo {volume} {67}},\ \bibinfo {pages} {022706} (\bibinfo {year} {2003})}\BibitemShut {NoStop}%
\bibitem [{\citenamefont {Storey}\ \emph {et~al.}(2016)\citenamefont {Storey}, \citenamefont {Pacholek}, \citenamefont {Levasseur}, \citenamefont {Dehning}, \citenamefont {Bodart}, \citenamefont {Rakai}, \citenamefont {Satou}, \citenamefont {Steyart}, \citenamefont {Sapinski},\ and\ \citenamefont {Schneider}}]{BGI_2016_development}%
  \BibitemOpen
  \bibfield  {author} {\bibinfo {author} {\bibfnamefont {J.}~\bibnamefont {Storey}}, \bibinfo {author} {\bibfnamefont {P.}~\bibnamefont {Pacholek}}, \bibinfo {author} {\bibfnamefont {S.}~\bibnamefont {Levasseur}}, \bibinfo {author} {\bibfnamefont {B.}~\bibnamefont {Dehning}}, \bibinfo {author} {\bibfnamefont {D.}~\bibnamefont {Bodart}}, \bibinfo {author} {\bibfnamefont {A.}~\bibnamefont {Rakai}}, \bibinfo {author} {\bibfnamefont {K.}~\bibnamefont {Satou}}, \bibinfo {author} {\bibfnamefont {D.}~\bibnamefont {Steyart}}, \bibinfo {author} {\bibfnamefont {M.}~\bibnamefont {Sapinski}},\ and\ \bibinfo {author} {\bibfnamefont {G.}~\bibnamefont {Schneider}},\ }\href@noop {} {\bibinfo {title} {Development of an ionization profile monitor based on a pixel detector for the cern proton synchrotron}} (\bibinfo {year} {2016}),\ \bibinfo {note} {dOI: 10.18429/JACoW-IBIC2015-TUPB059. Available at \url{https://cds.cern.ch/record/2265930}.}\BibitemShut {Stop}%
\bibitem [{\citenamefont {Storey}\ \emph {et~al.}(2017)\citenamefont {Storey}, \citenamefont {Bertsche}, \citenamefont {Bodart}, \citenamefont {Dehning}, \citenamefont {Gibson}, \citenamefont {Levasseur}, \citenamefont {Sandberg}, \citenamefont {Sapinski}, \citenamefont {Satou}, \citenamefont {Schneider} \emph {et~al.}}]{BGI_2017_first}%
  \BibitemOpen
  \bibfield  {author} {\bibinfo {author} {\bibfnamefont {J.}~\bibnamefont {Storey}}, \bibinfo {author} {\bibfnamefont {W.}~\bibnamefont {Bertsche}}, \bibinfo {author} {\bibfnamefont {D.}~\bibnamefont {Bodart}}, \bibinfo {author} {\bibfnamefont {B.}~\bibnamefont {Dehning}}, \bibinfo {author} {\bibfnamefont {S.}~\bibnamefont {Gibson}}, \bibinfo {author} {\bibfnamefont {S.}~\bibnamefont {Levasseur}}, \bibinfo {author} {\bibfnamefont {H.}~\bibnamefont {Sandberg}}, \bibinfo {author} {\bibfnamefont {M.}~\bibnamefont {Sapinski}}, \bibinfo {author} {\bibfnamefont {K.}~\bibnamefont {Satou}}, \bibinfo {author} {\bibfnamefont {G.}~\bibnamefont {Schneider}}, \emph {et~al.},\ }\href@noop {} {\bibinfo {title} {First results from the operation of a rest gas ionisation profile monitor based on a hybrid pixel detector}} (\bibinfo {year} {2017}),\ \bibinfo {note} {dOI:10.18429/JACoW-IBIC2017-WE2AB5. Proceedings of the 6th International Beam Instrumentation Conference (IBIC) 2017}\BibitemShut {NoStop}%
\bibitem [{\citenamefont {Ady}\ and\ \citenamefont {Kersevan}(2019)}]{molflowcitation}%
  \BibitemOpen
  \bibfield  {author} {\bibinfo {author} {\bibfnamefont {M.}~\bibnamefont {Ady}}\ and\ \bibinfo {author} {\bibfnamefont {R.}~\bibnamefont {Kersevan}},\ }\bibfield  {title} {\bibinfo {title} {Recent developments of {Monte-Carlo codes MolFlow+ and SynRad+}},\ }in\ \href {https://doi.org/10.18429/JACoW-IPAC2019-TUPMP037} {\emph {\bibinfo {booktitle} {Proceedings of the 10th International Particle Accelerator Conference (IPAC 2019)}}}\ (\bibinfo {address} {Melbourne, Australia},\ \bibinfo {year} {2019})\ p.\ \bibinfo {pages} {TUPMP037}\BibitemShut {NoStop}%
\bibitem [{\citenamefont {Ady}(2016)}]{molflow_algorithm_monte_carlo_ady2016}%
  \BibitemOpen
  \bibfield  {author} {\bibinfo {author} {\bibfnamefont {M.}~\bibnamefont {Ady}},\ }\emph {\bibinfo {title} {{Monte Carlo} simulations of ultra high vacuum and synchrotron radiation for particle accelerators}},\ \href@noop {} {Ph.D. thesis},\ \bibinfo  {school} {EPFL} (\bibinfo {year} {2016})\BibitemShut {NoStop}%
\bibitem [{\citenamefont {{Agilent Technologies}}(2016{\natexlab{a}})}]{agilent_vacion_300_2016}%
  \BibitemOpen
  \bibfield  {author} {\bibinfo {author} {\bibnamefont {{Agilent Technologies}}},\ }\href {https://www.agilent.com/cs/library/usermanuals/public/VacIon_Pump_300_Manual.pdf} {\emph {\bibinfo {title} {VacIon Plus 300 Pumps User Manual}}},\ \bibinfo {organization} {Agilent Technologies Italia S.p.A., Vacuum Products Division},\ \bibinfo {address} {Leinì (TO), Italy} (\bibinfo {year} {2016}{\natexlab{a}}),\ \bibinfo {note} {document No. 87-900-103-01 (H)}\BibitemShut {NoStop}%
\bibitem [{\citenamefont {{Agilent Technologies}}(2016{\natexlab{b}})}]{agilent_vacion_40_75_2016}%
  \BibitemOpen
  \bibfield  {author} {\bibinfo {author} {\bibnamefont {{Agilent Technologies}}},\ }\href {https://www.agilent.com/cs/library/usermanuals/public/VacIon_Pumps_40_55_75_Manual.pdf} {\emph {\bibinfo {title} {VacIon Plus 40/55/75 Pumps User Manual}}},\ \bibinfo {organization} {Agilent Technologies Italia S.p.A., Vacuum Products Division},\ \bibinfo {address} {Leinì (TO), Italy} (\bibinfo {year} {2016}{\natexlab{b}}),\ \bibinfo {note} {document No. 87-900-105-01 (G)}\BibitemShut {NoStop}%
\bibitem [{\citenamefont {Ku}\ \emph {et~al.}(1966)\citenamefont {Ku} \emph {et~al.}}]{error_propagation_formula}%
  \BibitemOpen
  \bibfield  {author} {\bibinfo {author} {\bibfnamefont {H.~H.}\ \bibnamefont {Ku}} \emph {et~al.},\ }\bibfield  {title} {\bibinfo {title} {Notes on the use of propagation of error formulas},\ }\bibfield  {journal} {\bibinfo  {journal} {Journal of Research of the National Bureau of Standards}\ }\textbf {\bibinfo {volume} {70}},\ \href {https://doi.org/https://dx.doi.org/10.6028/jres.070C.025} {https://dx.doi.org/10.6028/jres.070C.025} (\bibinfo {year} {1966})\BibitemShut {NoStop}%
\end{thebibliography}%
%\bibliography{main.bbl} % for arXiv

\end{document}